\documentclass[11pt,twoside]{article}
\usepackage[a4paper,hmargin=1.4in,vmargin=1.2in]{geometry}
\usepackage[utf8]{inputenc}
\usepackage[T1]{fontenc}
\usepackage[english]{babel}
\usepackage[labelfont=bf]{caption}
\usepackage{subcaption}
\usepackage{array}
\usepackage[numbers]{natbib}
\usepackage{multirow}
\usepackage{booktabs}
\usepackage{amsmath}
\usepackage{amssymb}
\usepackage{amsthm}
\usepackage{mathrsfs}
\usepackage{mathtools}
\usepackage{stackrel}
\usepackage{bbm}
\usepackage{enumerate}
\usepackage{url}
\usepackage{xcolor}
\usepackage{hyperref}
\usepackage{algpseudocode}
\usepackage{algorithmicx}
\usepackage{algorithm}
\usepackage{csquotes}

\newtheorem{theorem}{Theorem}
\newtheorem{proposition}{Proposition}

\newtheorem{lemma}{Lemma}

\numberwithin{equation}{section}
\numberwithin{theorem}{section}
\numberwithin{lemma}{section}
\numberwithin{proposition}{section}

\newtheorem{remark}{Remark}

\DeclareMathAlphabet{\mathpzc}{OT1}{pzc}{m}{it}
\DeclareMathOperator*{\argmin}{arg\,min}

\newcommand{\ES}{\mathbb{E}\mathbb{S}}
\newcommand{\VaR}{\mathbb{V}\mathrm{a}\mathbb{R}}

\DeclareMathOperator*{\Argmin}{Argmin}

\title{Mirror Descent Algorithms for Risk Budgeting Portfolios}

\author{Martin~Arnaiz~Iglesias \thanks{Universit\'e Paris 1 Panth\'eon-Sorbonne, Centre d'Economie de la Sorbonne (CES), 106 Boulevard de l’H\^opital, 75642 Paris
Cedex 13 ({\tt Martin.Arnaiz-Iglesias@etu.univ-paris1.fr}).}
\and
Adil~Rengim~Cetingoz\thanks{Universit\'e Paris 1 Panth\'eon-Sorbonne, Centre d'Economie de la Sorbonne (CES), 106 Boulevard de l’H\^opital, 75642 Paris
Cedex 13 ({\tt rengimcetingoz@gmail.com}).}
\and
Noufel~Frikha\thanks{Universit\'e Paris 1 Panth\'eon-Sorbonne, Centre d'Economie de la Sorbonne (CES), 106 Boulevard de l’H\^opital, 75642 Paris
Cedex 13 ({\tt noufel.frikha@univ-paris1.fr}). The research of N. Frikha has benefited from the support of the Institut Europlace de Finance.}
}

\date{\today}

\newcommand{\beqa}{\begin{eqnarray}}
\newcommand{\eeqa}{\end{eqnarray}}
\def\bal{\begin{aligned}}
\def\eal{\end{aligned}}

\hypersetup{
 colorlinks  = true, %
 urlcolor   = black, %
 linkcolor  = black, %
 citecolor  = black %
}
\makeatletter
\renewcommand\@makefnmark{\hbox{\@textsuperscript{\normalfont\color{blue}\@thefnmark}}}
\renewcommand\@makefntext[1]{%
 \parindent 1em\noindent
      \hb@xt@1.8em{%
        \hss\@textsuperscript{\normalfont\@thefnmark}}#1}
\makeatother

\usepackage{xcolor}

\begin{document}

\maketitle

\begin{abstract}
This paper introduces and examines numerical approximation schemes for computing risk budgeting portfolios associated to positive homogeneous and sub-additive risk measures. We employ Mirror Descent algorithms to determine the optimal risk budgeting weights in both deterministic and stochastic settings, establishing convergence along with an explicit non-asymptotic quantitative rate for the averaged algorithm. A comprehensive numerical analysis follows, illustrating our theoretical findings across various risk measures -- including standard deviation, Expected Shortfall, deviation measures, and Variantiles -- and comparing the performance with that of the standard stochastic gradient descent method recently proposed in the literature.
\paragraph{Keywords:} Risk Budgeting, risk measures, Mirror Descent, Monte Carlo, numerical finance.

\medskip
\paragraph{MSC:} 65C05, 62L20, 62G32, 91Gxx.
%	65C05  	Monte Carlo methods
% 	62L20  	Stochastic approximation
% 91G60  	Numerical methods (including Monte Carlo methods)
% 91Gxx		Mathematical finance
%	62G32  	Statistics of extreme values; tail inference 
\end{abstract}

\section{Introduction}
The financial problem of constructing investment portfolios has often been analyzed as a mathematical optimization problem, at least since Markowitz \cite{markowitz1952} formulated it as maximizing a portfolio’s expected return under a constraint on risk, defined by variance. Solving this problem to obtain the optimal portfolio typically requires numerical methods, except in trivial cases where analytical solutions exist, such as when the only constraint is that portfolio weights sum to one. Today, efficient tools exist to compute this portfolio under almost any set of convex constraints, given the expected return vector and the covariance matrix of asset returns (see \cite{markowitz1956optimization}, \cite{wolfe1959simplex}, and \cite{boyd2024markowitz}). However, this framework is rarely adopted in its classical form in real-life applications, mainly due to the fact that it heavily depends on accurate estimates of expected returns, variances, and covariances. Small errors in these inputs, especially in expected returns, can lead to vastly different and potentially suboptimal portfolio allocations (\cite{best1991sensitivity} and \cite{michaud1989markowitz}). 

Alternative portfolio construction approaches that prioritise risk management and diversification over return maximisation have thus been proposed -- especially following the dot-com crash and the Great Financial Crisis -- to address investment objectives beyond optimising a mean-variance criterion. Risk budgeting is one such approach, aiming to construct a long-only portfolio in which each asset contributes to the portfolio's total risk according to a risk budget specified by the investor. For a comprehensive introduction, we refer to \cite{roncalli2013introduction}. This method is thus closely aligned with the principles of diversification, helping to mitigate concentrated risks within the portfolio. 

The risk budgeting problem is mathematically formulated as a system of nonlinear equations, which also corresponds to the first-order conditions of a constrained convex minimisation problem. Leveraging this variational characterisation, and under positive homogeneous and sub-additive risk measures, the existence of a unique solution has been extensively studied in the literature; see, for example, \cite{BR:2012} and \cite{cetingoz2023risk}. However, an analytical solution is generally not obtainable, prompting the development of various numerical methods for computing risk budgeting portfolios.

Volatility has been one of the earliest and most extensively studied risk measures in the context of risk budgeting \cite{qian2005risk}. Initial efforts, including those by \cite{maillard2010properties} and \cite{bai2016least}, focused on minimizing a least-squares formulation. Subsequently, \cite{chaves2012efficient} proposed employing Newton's method, with \cite{S:2013} further advancing this approach by demonstrating the feasibility of Nesterov acceleration, leveraging the self-concordant nature of the objective function to achieve provable convergence. Additionally, \cite{GBRR:2013} advocates for a cyclical coordinate descent algorithm, which proves particularly efficient for high-dimensional portfolios. 

Risk measures beyond volatility have also attracted interest in risk budgeting. Expected Shortfall (ES for short) is one of the most extensively studied, as it is considered to be more representative of the true risk of financial investments and satisfies certain desirable mathematical properties, as outlined in \cite{AT02} and \cite{rockafellar2002conditional}. For example, \cite{mausser2018long} proposes formulating the problem as a second-order cone program when asset returns are modeled by discrete distributions. For continuous distributions, \cite{da2023risk} introduces a cutting planes algorithm, while \cite{city33733} develops a numerical procedure based on a novel ES estimator that also facilitates statistical inference of ES-based risk budgeting portfolios. Other risk measures have also been considered, such as Mean Absolute Deviation (MAD) \cite{ararat2024mad} and Expectiles \cite{bellini2021risk}, highlighting the importance of flexibility in defining risk using a diverse set of risk measures in the risk budgeting framework.

A comprehensive stochastic formulation of the risk budgeting problem is presented in \cite{cetingoz2023risk}, grounded in a unified probabilistic framework applicable to a wide range of risk measures. Specifically, the problem is cast as a stochastic optimisation task whenever the risk measure can be represented as the minimum of a convex function expressed in expectation form—a structure that encompasses, for instance, Bayes risk measures \cite{bayes:risks:21}. To enable the computation of risk budgeting portfolios across an extensive class of risk measures—including entire families of spectral risk measures \cite{acerbi2002spectral} and deviation measures as defined in \cite{rock:uryasev:zaba:06}—the authors of \cite{cetingoz2023risk} propose employing a standard projected stochastic gradient descent (SGD) algorithm, though its convergence properties are not examined.

Our primary objective in this paper is to propose and study Mirror Descent (MD) algorithms for computing risk budgeting portfolios associated with general positive homogeneous and sub-additive risk measures. Originating from the pioneering work of Nemirovskij and Yudin \cite{nemirovskij1983problem}, MD provides a natural framework for addressing optimisation problems, particularly when the mirror mapping (also known as proximal mapping) is explicit. MD has the appealing property of defining a smooth trajectory that remains within the constrained set, thus eliminating the need for additional projection steps, as required in the SGD algorithm proposed in \cite{cetingoz2023risk}. This effectively pushes the boundaries of the domain to an infinite distance from any interior point. For a comprehensive introduction and extensions to the stochastic setting and convex-concave stochastic saddle-point problems, we refer to \cite{nesterov:07} and \cite{NJLS09}. Additional almost sure convergence results for convex and non-convex stochastic optimisation problems can be found in \cite{luan:Nemirovski:Shapiro} and \cite{NIPS2017_e6ba70fc}. We also refer to the recent work \cite{costa:gadat:huang} in which a stochastic MD scheme is proposed for optimal portfolio allocation with ES penalty. Thus, MD appears to be a natural method to implement for solving the constrained convex minimisation problem associated with the risk budgeting problem.

However, a key assumption for deriving $L^1(\mathbb{P})$ convergence (also known as regret bounds) and concentration inequalities for MD schemes -- the boundedness of the objective's gradient -- is not satisfied in our context. Indeed, the gradient of the objective function diverges on the boundary of the domain, preventing direct application of standard results (see, e.g. \cite{NJLS09}). To overcome this major challenge, we instead employ a tailored, tamed version of the gradient, which, like the original gradient, vanishes only at the unique minimiser of the objective function but remains uniformly bounded across the entire domain. This modified gradient enables us to develop both deterministic and stochastic MD schemes and to establish their convergence to the unnormalised risk budgeting portfolio for general risk measures under mild assumptions. In both deterministic and stochastic settings, we further establish a non-asymptotic $a.s.$ convergence rate for the weighted averaged sequence. Our stochastic framework accommodates a wide range of risk measures, including volatility, ES, deviation measures, and Variantiles, among others. To the best of our knowledge, our scheme is the first to ensure $a.s.$ convergence along with a non-asymptotic quantitative convergence rate for computing risk budgeting portfolios across general risk measures.

The paper is organized as follows. In Section~\ref{sec:riskbudgeting}, we briefly introduce the risk budgeting problem, emphasising its formulation as the solution to a strictly convex optimisation problem, which serves as the cornerstone for our analysis in the subsequent sections. Section~\ref{sec:mirror_descent} begins by presenting our tailored, tamed version of the gradient and some key results essential for establishing the convergence of our schemes. We then introduce our deterministic MD (DMD) algorithm, providing convergence results along with a convergence rate for the weighted averaged sequence. Next, we propose and analyse a stochastic MD (SMD) scheme in a stochastic framework where the risk measure is expressed as the minimum of a convex map. We establish the $a.s.$ convergence of the SMD algorithm, along with a non-asymptotic $a.s.$ convergence rate for the weighted average sequence. In Section~\ref{section:numerics}, we conduct an extensive numerical analysis to assess the practical advantages of the MD approach over standard projected gradient descent methods, particularly in the stochastic setting, for accurately computing risk budgeting portfolios of various sizes under different risk measures. The proofs of the main and auxiliary results are deferred to Appendix~\ref{sec:appendix}.

\noindent \textbf{Notations:} $\mathbb{R}^0$ stands for $\mathbb{R}\backslash{\left\{0\right\}}$ and $\mathbb{R}^0_+$ for $\mathbb{R}_+\backslash{\left\{0\right\}}$. For a vector $x \in \mathbb{R}^d$, we write $\|x\|=\|x\|_1= \sum_{i=1}^d |x_i|$ for its norm and $\|x\|_{\infty}= \max_{1\leq i\leq d}|x_i| =  \sup_{\|y\|_1\leq 1}\langle y, x \rangle$ stands for the dual norm of $\|.\|_1$. The simplex of dimension $d$ is denoted by $\Delta_d= \left\{ u \in \mathbb{R}_+^{d} : u_1 + \cdots + u_d = 1\right\}$ and we let let $\Delta_d^{>0} := \left\{ u \in (\mathbb{R}_+^{0})^d: u_1+\cdots + u_d = 1\right\}$.

\section{The Risk Budgeting Problem}
\label{sec:riskbudgeting}
We consider a financial market composed of $d$ assets whose returns are given by the $\mathbb{R}^d$-valued random variable $X$ on a probability space $(\Omega, \mathcal{F}, \mathbb{P})$ which is assumed to be rich enough to accommodate all of the subsequent random variables.
A financial portfolio is identified with its corresponding vector of weights $u = (u_1, \cdots, u_d)$ belonging to the simplex $\Delta_d$. If the portfolio is given by the vector $u \in \Delta_d$, then $-\langle u, X\rangle = -\sum_{i=1}^d u_i X_i$ corresponds to its loss, recalling that $\langle .,.\rangle$ stands for the Euclidean scalar product on $\mathbb{R}^d$.

In order to assess the risk of the loss of a financial portfolio, we consider a risk measure $\rho$ that is a function mapping a random variable $Z$ to the real number $\rho(Z)\in \mathbb{R}$. We deliberately omit the space on which $\rho$ is defined but typically one has $Z \in L^{0}(\mathbb{P}), \, L^{1}(\mathbb{P})$ or $L^{2}(\mathbb{P})$.

The risk measure $\rho$ is said to be risk budgeting compatible (RB-compatible for short) if it is positive homogeneous and sub-additive, namely if for any $\lambda \geq 0$, 
\begin{equation}\tag{PH}
\rho(\lambda Z) = \lambda \rho(Z) 
\end{equation}
\noindent and, for any real-valued random variables $Z_1$ and $Z_2$
\begin{equation}\tag{SA}
\rho(Z_1 + Z_2) \leq \rho(Z_1) + \rho(Z_2).
\end{equation}

To any RB-compatible risk measure $\rho$, we associate the map 
$$
r_\rho :  \mathbb{R}_+^d \ni y \rightarrow \rho(-\langle y, X \rangle) \in \mathbb{R}.
$$

In particular, the risk associated to the portfolio $u$ is thus given by $r_\rho(u)=\rho(-\langle u, X\rangle)$.

In what follows, we will always assume that the map $r_\rho$ is continuous on $\mathbb{R}^d_+$ and continuously differentiable on the open set $(\mathbb{R}^{0}_+)^d$. The positive homogeneity property of the risk measure $\rho$ combined with Euler’s homogeneous function theorem implies that the risk linked with a portfolio represented by its vector of weights $u$ can be decomposed as follows 
\begin{equation}
\label{decomposition:risk:euler:thm}
r_\rho(u) = \sum_{i=1}^d u_i \partial_{u_i} r_\rho(u), \quad u \in \Delta^{>0}_d.
\end{equation}

In the risk budgeting literature, the $i$-th term of the above sum, $u_i \partial_{u_i} r_\rho(u)$, is referred to as the risk contribution of asset $i$ to the overall portfolio risk. The risk budgeting problem, therefore, involves identifying a vector of weights $u^{\star}$ such that the risk contributions align with predetermined proportions, represented by a vector of risk budgets $b$, of the total risk.

To be more specific, for a given vector of risk budget $b \in \Delta_d^{>0} $, we say that $u \in \Delta_d^{>0}$ solves the risk budgeting problem $RB_b$ if the following condition holds:
$$
u_i \partial_{u_i} r_\rho(u) = b_i r_\rho(u), \quad i=1, \cdots, d.
$$

Throughout the paper, we will assume that the risk of any long-only portfolio is positive, that is, $r_\rho(u)>0$, for any $u\in \Delta_d$. We refer to \cite{cetingoz2023risk} and also \cite{bardou2016cvar,frikha:sifin:1} where a similar assumption is made in the context of ES and shortfall risk minimization.

The existence and uniqueness of a vector of weights that solves $RB_b$ for a given vector of risk budgets $b \in \Delta_d^{>0}$, along with its characterization as the solution to a strictly convex optimization problem, has been investigated in several works, see for instance \cite{city33733,cetingoz2023risk,BR:2012,GBRR:2013,S:2013}. Here, we recall the characterization provided in \cite{cetingoz2023risk}.

\begin{theorem}\label{thm:cetingoz:fermanian:gueant}{\cite[Theorems 1 and 2]{cetingoz2023risk}}
Let $b \in \Delta_d^{>0}$. Let $g: \mathbb{R}_+ \rightarrow \mathbb{R}$ be a continuously differentiable, convex and increasing function. Let the function $\Gamma_g : (\mathbb{R}^{0}_+)^d \rightarrow \mathbb{R}$ be defined by
$$
\Gamma_g: y \mapsto g(r_\rho(y)) - \sum_{i=1}^{d} b_i \log y_i.
$$

 There exists a unique minimizer $y^{\star}$ of the strictly convex function $\Gamma_g$ satisfying $\nabla \Gamma_g(y^\star) = 0$ and 
 $$
 u^{*} := \frac{y^\star}{\|y^{\star}\|_1}
 $$
 \noindent solves the risk budgeting problem $RB_b$. Moreover, if $u$ is a solution to $RB_b$ then 
 $$
 u = u^\star.
 $$
\end{theorem}

We highlight that this result serves as the cornerstone of our analysis. Indeed, the numerical schemes we propose will be specifically designed to target the unique minimizer $y^{\star}$ of the aforementioned strictly convex optimization problem. Theorem \ref{thm:cetingoz:fermanian:gueant} subsequently demonstrates that the renormalized minimizer yields the unique solution to the risk budgeting problem $RB_b$.

\section{Mirror Descent Algorithm}
\label{sec:mirror_descent}
To solve the minimization problem introduced in Theorem \ref{thm:cetingoz:fermanian:gueant}, we are naturally led to use gradient descent schemes and their stochastic counterpart when the map $\rho$ writes as an expectation of a random function. However, several specific difficulties appear. The first one is the fact that the gradient of $\Gamma_g$ diverges on $\partial (\mathbb{R}_+^{0})^d$ due to the presence of the term $\sum_{i=1}^d b_i \log(y_i)$, thus requiring a specific treatment. The second difficulty is that the minimization problem is set over $(\mathbb{R}_+^0)^d$ and not $\mathbb{R}^d$. 

To address the first challenge, we will employ a tailored, tamed version of the gradient, which, like $\nabla \Gamma_g$, vanishes only at the unique minimizer of $\Gamma_g$. The second challenge, stemming from the constrained nature of the minimisation problem, is managed using the MD algorithm-a flexible optimisation technique widely used in convex optimisation and machine learning, as noted in the introduction. This algorithm has the advantageous property of defining a trajectory that remains entirely within the constrained set, thereby eliminating the need for additional projection steps that could be difficult to implement in practice. 

\subsection{Taming the singularity of the gradient}\label{sub:sec:taming:singularity} 

The target of our MD algorithm is the unique minimizer $y^{\star}$ of the strictly convex map $\Gamma_g$. However, as already mentioned, a significant challenge in efficiently implementing this scheme lies in the divergence of the gradient at the boundary of its domain $\partial (\mathbb{R}_+^{0})^d$. Specifically, since $b_i>0$, it follows that $\partial_{y_i}  \Gamma_g(y) \rightarrow -\infty$, as soon as $y_i\downarrow 0$ for some $i \in \left\{1, \cdots, d\right\}$. Hence, $\sup_{y \in (\mathbb{R}^0_+)^d}\|\nabla \Gamma_g(y)\|_\infty = \infty$.  

As highlighted in \cite{NJLS09}, the boundedness of the objective function's gradient on the constrained set is a crucial prerequisite for establishing a quantitative convergence rate for this type of recursive procedure.

%We will make the following assumption polynomial growth assumption
%\begin{assumption}\label{assump:polynomial:growth}
%There exists an integer $p$ and a constant $C<\infty$ such that for any $y \in \mathbb{R}_+^d$, 
%$$
%\|\nabla (g(\rho(-\langle y, X\rangle)))\| = g'(\rho(-\langle y,X\rangle)) \|\nabla (\rho(-\langle y, X \rangle)) \|_{\infty}  \leq C (1+\|y\|_\infty^p).
%$$
%\end{assumption}
%\end{comment}

In order to circumvent this major difficulty, we build our numerical procedure on a tamed version of $\nabla \Gamma_g$. We let $\underline{y}= \min_{1\leq i \leq d} y_i$ and define
$$
\kappa(y) = \underline{y}\wedge 1, \quad y \in \mathbb{R}_+^d.
$$

We then remark that the map $(\mathbb{R}^0_+)^d \ni y \mapsto \kappa(y) \, \nabla  \Gamma_g(y)$ can be extended by continuity on $\mathbb{R}_+^d$.\footnote{Throughout the article, we will use the convention $0/0 = 1$.} Indeed, note that if $y_i=0$ for some $i \in \left\{1, \cdots, d\right\}$, then, $\underline{y} = 0$ so that 
\begin{align*}
(\kappa(y) \, \nabla \Gamma_g(y))_j & = \underline{y} \, g'(r_\rho(y)) \partial_{y_j} r_\rho(y)  -  b_j \frac{\underline{y}}{y_j} \\
& = 
\begin{cases}
- b_j, &  \mbox{ if } y_j=\underline{y}=0 \\
 0, & \mbox{otherwise} 
\end{cases} 
\end{align*}
\noindent so that $\kappa(y) \, \nabla \Gamma_g(y) \neq 0$ on $\mathbb{R}_+^d \backslash (\mathbb{R}^{0}_+)^d$. We thus conclude that the unique minimizer $y^{\star}$ of $\Gamma_g$ satisfies
$$
\left\{ y^{*} \right\} = \left\{ y \in (\mathbb{R}^0_+)^d : \nabla  \Gamma_g(y) = 0 \right\}  = \left\{ y \in \mathbb{R}_+^d : \kappa(y) \, \nabla \Gamma_g(y) = 0 \right\}.
$$

It is here important to emphasize that, in contrast to  $\nabla \Gamma_g$, the map $y \mapsto \kappa(y) \, \nabla \Gamma_g(y)$  remains bounded on any centred closed ball $B_m$ of radius $m$ and dimension $d$, since for any $y\in B_m$, the following holds
\begin{equation}\label{ineq:growth:modified:grad}
\begin{aligned}
\|& \kappa(y)\nabla  \Gamma_g(y)\|_\infty \\
& \leq  \max_{y\in B_m}\left\{g'(r_\rho(y)) \|\nabla r_\rho(y) \|_\infty\right\}  + \max_{1\leq i\leq d} b_i \frac{\underline{y}\wedge 1}{y_i} \\
& \leq  M_\star(m, b, d) := \max_{y\in B_m}\left\{g'(r_\rho(y)) \|\nabla r_\rho(y) \|_\infty\right\}+ \max_{1\leq i \leq d} b_i .
\end{aligned}
\end{equation}

We conclude this section by the following technical lemma that will play a central role to address the convergence of our algorithm.
\begin{lemma}\label{lem:mean:reverting} For all $y \in \mathbb{R}_+^d$, $y\neq y^{\star}$, it holds 
$$
\langle y - y^{\star}, \kappa(y) \nabla \Gamma_g(y) \rangle >0.
$$
\end{lemma}
\begin{proof}
    The strict convexity of $\Gamma_g$ guarantees that the desired property holds on $(\mathbb{R}_+^{0})^d$. Now, if $y \in \mathbb{R}_+^d \backslash (\mathbb{R}_+^{0})^d$ then $\underline{y}= 0$ so that
    $$
\langle y - y^{\star}, \kappa(y) \nabla \Gamma_g(y) \rangle =  \sum_{j: y_j =0} (y_j - y^{\star}_j) (-b_j) =  \sum_{j: y_j =0} y^{\star}_j b_j >0
    $$
\noindent which concludes the proof.
\end{proof}

\subsection{Deterministic Mirror Descent Algorithm}\label{deterministic:MD}

Although our primary objective is to address the stochastic framework, where $\rho$ is expressed as an expectation, in certain practical settings the function  $y\mapsto r_\rho(y)=\rho(-\langle y, X\rangle)$ is known explicitly or semi-explicitly. In such cases, it is advantageous to rely on a deterministic Mirror Descent (DMD) algorithm to approximate $y^\star$. We introduce the Bregman divergence 
$$
D_F(y ,y') = F(y) - F(y') - \langle \nabla F(y') , y-y' \rangle, \quad y, y' \in (\mathbb{R}_+)^d
$$
\noindent generated by the negative entropy function $F(y) = \sum_{i=1}^d y_i \log(y_i)$, $y\in \mathbb{R}_+^d$. The map $D_F$ is also known as the Kullback–Leibler divergence inasmuch it satisfies
$$
D_F(y, y')=\sum_{i=1}^d y_i \log(y_i/y'_i) - \sum_{i=1}^d y_i + \sum_{i=1}^d y_i'. 
$$

Let $m>0$. Using the fact that the map $[0, m] \ni y\mapsto y\log(y) - \frac{1}{2 m} y^2$ is convex, one deduces
\begin{equation}\label{ineq:strong:convexity}
D_F(y, y') \geq \frac{1}{2 m} \|y-y'\|^2_1, \quad \mbox{for any } y, y'  \in \mathbb{R}_+^d \cap B_m.
\end{equation}

Let $(\gamma_k)_{k\geq1}$ be a deterministic positive learning sequence. Starting from $y^0 \in (\mathbb{R}^{0}_+)^d \cap B_m$, the deterministic MD algorithm constructs the sequence $(y^k)_{k\geq1}$ inductively where at step $k$, given $y^k \in (\mathbb{R}_+^{0})^{d}\cap B_m$, $y^{k+1}$ is obtained by solving the optimization problem
$$
y^{k+1} = {\rm P}^m_{y^k}(\gamma_{k+1} \kappa(y^k) \nabla  \Gamma_g(y^k))
$$

\noindent where, for a given $y \in (\mathbb{R}^{0}_+)^d \cap B_m$, the proximal mapping ${\rm P}^{m}_y: \mathbb{R}^d \rightarrow (\mathbb{R}^{0}_+)^{d}\cap B_m$, associated with $D_F$, is defined by
\begin{equation}\label{def:proximal:map}
{\rm P}^{m}_y(v)=\argmin_{w \in (\mathbb{R}^{0}_+)^{d} \cap B_m}\left\{  \langle  v, w-y \rangle + D_F(y, w)\right\} =   \left\{
                \begin{array}{ll}
                  y e^{-v} ,  \mbox{ if } \| y e^{-v}\|_1 \leq m,\\
\\                
                   \frac{m}{\| y e^{-v}\|_1} \, y e^{-v}, \, \mbox{ otherwise, }
                \end{array}
              \right. 
\end{equation}

\noindent where $ye^{-v} = (y_1 e^{-v_1}, \cdots, y_de^{-v_d})$.

The key point here is that the unique solution to the above constrained strictly convex minimization problem is explicit. Indeed, from \eqref{def:proximal:map}, it directly follows that $y^{k+1}=(y^{k+1}_1, \cdots, y^{k+1}_d)$, for $k\geq0$, is given by
\begin{equation}\label{rec:deterministic:md}
\begin{aligned}
     y^{k+1}_i & = ( {\rm P}^m_{y^k}(\gamma_{k+1} \kappa(y^k) \nabla  \Gamma_g(y^k)) )_i \\
    &  = \left\{
                \begin{array}{ll}
                  y^k_i e^{-\gamma_{k+1} \kappa(y^k)\, \partial_{y_i}  \Gamma_g(y^k)} ,  \mbox{ if } \| y^k e^{-\gamma_{k+1} \kappa(y^k) \nabla  \Gamma_g(y^k)}\|_1 \leq m,\\
\\                
                   \frac{m}{\| y^k e^{-\gamma_{k+1} \kappa(y^k) \nabla  \Gamma_g(y^k)}\|_1 } \, y^k_i e^{-\gamma_{k+1} \kappa(y^k)) \partial_{y_i}  \Gamma_g(y^k)}, \, \mbox{ otherwise }
                \end{array}
              \right. \quad i=1, \cdots, d.
\end{aligned}
\end{equation}

The recursive scheme outlined above can be implemented as soon as the gradient $\nabla \Gamma_g$ is available. The following theorem presents the principal result of this section, with its proof deferred to Appendix \ref{proof:thm:conv:DMD}.

\begin{theorem}\label{thm:conv:DMD} Assume that $m\geq \|y^{\star}\|_1$. Assume additionally that the learning step sequence $(\gamma_n)_{n\geq1}$ satisfies $\sum_{n\geq1}\gamma_n=\infty$ and $\sum_{n\geq1} \gamma_n^2< \infty$. Then, the sequence $(y^n)_{n\geq0}$ defined by \eqref{rec:deterministic:md} converges to $y^{\star}$ as $n\uparrow \infty$. Moreover, the weighted averaged sequence $(\bar{y}^{n})_{n\geq 1}$ of $(y^n)_{n \geq 0}$ defined by
$$
\bar{y}^n= \frac{\sum^n_{k=1} \gamma_k y^{k-1}}{\sum^n_{k=1} \gamma_k}, \quad n\geq 1,
$$
\noindent satisfies
\begin{equation}\label{conv:rate:deterministic:md}
\Gamma_g(\bar{y}^{n}) - \Gamma_g(y^{\star}) \leq \frac{D_F(y^\star,y^0) + \frac12 m M^2_{\star} \sum^n_{k=0} \gamma_{k+1}^2}{ (\min_{0\leq k\leq n-1} \underline{y}^k \wedge 1) \sum^n_{k=0} \gamma_{k+1}}, \quad n\geq1,
\end{equation}

\noindent where $M_\star$ is given by \eqref{ineq:growth:modified:grad}.
\end{theorem}

\begin{remark}\label{remark:convergence:rate}
Concerning the convergence rate given by \eqref{conv:rate:deterministic:md}, under the two conditions $\sum_{n\geq1}\gamma_n = \infty$ and $\sum_{n\geq1} \gamma_n^2< \infty$, one may select $\gamma_n = \gamma n^{-\frac12 - \delta}$, with $\delta \in (0, 1/2]$ and $\gamma>0$. A straightforward comparison between series and integrals then yields 
$$
\Gamma_g(\bar{y}^{n}) - \Gamma_g(y^{\star}) \leq K n^{-\frac12 + \delta}, \quad n\geq1.
$$

 Alternatively, choosing $\gamma_n = n^{-1/2} \log(n)^{ -(\frac12 + \delta)}$, for some $\delta >0$, ensures that the Bertrand series $\sum_{n\geq1} \gamma_n^2$ converges, yielding
$$
\Gamma_g(\bar{y}^{n}) - \Gamma_g(y^{\star}) \leq K \frac{(\log{n})^{\frac12+\delta} }{ \sqrt{n} }, \quad n\geq1,
$$

\noindent for some constant $K< \infty$ which depends on $m$, $d$, $M_\star$ and $\delta$. Thus, we recover the usual convergence rate achieved by deterministic MD schemes, up to a logarithmic factor \cite{NJLS09}.
\end{remark}

\begin{remark}\label{remark:initialisation:algo} An explicit bound for the initialisation error term $D_F(y^\star, y^0)$ can be established in terms of the parameter $m$ and the dimension $d$ by appropriately choosing $y^0$. Specifically, if $y^0$ is selected as $\argmin_{\mathbb{R}_+^d \cap B_m} F= e^{-1}(1, \cdots, 1)$ if $m\geq e^{-1}d$ or $m d^{-1}(1,\cdots, 1)$ otherwise, then $F(y^0)=\min_{\mathbb{R}_+^d \cap B_m} F \geq -d e^{-1}$. Furthermore, if $m\leq 1$ then since $F\leq 0$ on $\mathbb{R}_+^d \cap B_m$, we have $\max_{\mathbb{R}_+^d \cap B_m} F\leq 0$; whereas, if $m\geq1$, then $F(x)=\|x\|_1 \sum_{i=1}^d \left\{ \frac{x_i}{\|x\|_1}\log(\frac{x_i}{\|x\|_1}) + \frac{x_i}{\|x\|_1}\log(\|x\|_1) \right\} \leq m\log(m)$ for any $x\in \mathbb{R}_+^d \cap B_m$. Hence, 
$$
D_F(y^{\star},y^{0}) \leq \max_{\mathbb{R}_+^d \cap B_m} F - \min_{\mathbb{R}_+^d \cap B_m} F \leq (m\log(m))_+ +\frac{d}{e}. 
$$

This upper bound is particularly valuable as it illuminates the dependence of the convergence rate on the dimension, indicating that this rate grows at most linearly with $d$.   
\end{remark}

\begin{remark}\label{remark:selection:m}
    The above result indicates that one must select $m$ sufficiently large, ensuring that $m\geq \|y^{\star}\|_1$. Given that $y^\star$ is unknown, this necessitates a somewhat arbitrary choice for the user.\footnote{In fact, one can use the identity $r_\rho(y^\star) = \frac{1}{g'(r_\rho(y^\star))}$ (from Theorem~\ref{thm:cetingoz:fermanian:gueant}) to determine a reasonable choice for $m \geq \|y^\star\|_1$. For example, when $g = \text{Id}$, this implies $r_\rho(y^\star) = 1$, indicating that the risk of the portfolio under the weights $y^\star$ equals 1. This equality provides insight into the order of magnitude of the elements of $y^\star$.} From a numerical standpoint, it is evident that choosing $m$ too small -- such that $m< \|y^\star\|_1$ -- will prevent the algorithm from converging to $y^\star$. With respect to the convergence rate, although the upper bound is indeed influenced by selecting an excessively large value of $m$, we do not observe a significant impact on the convergence rate of the weighted averaged sequence $(\bar{y}^n)_{n\geq1}$ during practical implementations. We refer the reader to Section \ref{section:numerics} for a discussion of numerical evidence.
\end{remark}

\subsection{Stochastic Mirror Descent}

\subsubsection{Stochastic framework}\label{subsubsec:stochastic:framework}

In this section, we propose and analyze a stochastic version of the previous MD algorithm within a stochastic framework. This is achieved in the presence of a convex loss function $\mathbb{R}^2 \ni  (\xi, x) \mapsto L(\xi, x)$, such that $L(\xi, -\langle y, X\rangle) \in L^{1}(\mathbb{P})$ for any $(\xi, y) \in \mathbb{R}\times \mathbb{R}^d_+$, and satisfying the condition:
\begin{equation}\label{def:g:r:rho:stochastic:case}
g(r_\rho(y)) = \min_{\xi \in \mathbb{R}} \mathbb{E}[L(\xi, -\langle y, X\rangle)] = \mathbb{E}[L(\xi^\star(y), -\langle y, X\rangle)], \quad y \in (\mathbb{R}^0_+)^d,
\end{equation}

\noindent  where the minimizer $\xi^\star(y)$ is assumed to be uniquely defined for every $y\in (\mathbb{R}^0_+)^d$.

As already noticed in \cite{cetingoz2023risk}, when $g=\mbox{Id}$, the associated risk measure is linked to the notions of Bayes pair and Bayes risk measure mathematically characterized as the minimization of an expected loss function over a set of possible decisions \cite{bayes:risks:21}. Such risk measures ensure robust risk minimization and optimal decision-making under uncertainty. 

From both practical and theoretical perspectives, it is valuable to consider the general framework where $g\neq \mbox{Id}$. Perhaps the most significant example is volatility, which attains its minimum under the square map $g(x)=x^2$.While a stochastic algorithm may not be the most efficient numerical method for addressing the problem when volatility is chosen as the risk measure, it is crucial that the stochastic framework accommodates volatility, as it is the most widely used risk measure in the risk budgeting paradigm. Other examples include $L^{p}$-deviation risk measures for $p\geq1$, which similarly are characterized by a minimum under the map $g(x)=x^p$. For a more detailed discussion, we refer to Section \ref{subsection:examples}. 

%Given a random variable Z representing a financial position or loss, and a convex loss function , the convex Bayes risk measure is defined by
%$$
%\rho(Z) = \min_{\xi \in \mathbb{R}} \mathbb{E}[L(\xi, Z)].
%$$

%\noindent where for some positive real numbers $a, \, b$,
%$$
%\psi_{a, b}(z) = a z_+ + b z_-, \quad z \in \mathbb{R}.
%$$
In this framework, the optimization problem related to the RB problem equivalently writes
\begin{equation}\label{reformulation:rb:problem:bayes:risk}
\min_{y \in (\mathbb{R}^{0}_+)^d} \Gamma_g(y) =\min_{z:=(\xi, y) \in \mathbb{R} \times (\mathbb{R}^{0}_+)^d} \left\{ h(z) := \mathbb{E}\left[H(z, X)\right]  \right\}
\end{equation}
\noindent where
$$
 H(z, X) := L(\xi, -\langle y,X\rangle ) - \sum_{i=1}^d b_i \log(y_i), \quad z=(\xi, y) \in \mathbb{R} \times (\mathbb{R}^{0}_+)^d.
$$

The following result demonstrates the existence of a solution to the above optimization problem, which, in turn, provides the unique solution to the risk budgeting problem. The proof of this result is deferred to Appendix \ref{subsec:prop:rb:problem:stochastic:framework}.

\begin{proposition}\label{prop:rb:problem:stochastic:framework} Assume that the function $(\xi , x)\mapsto L(\xi, x)$ satisfies the following regularity and integrability conditions: 
\begin{itemize}
\item $\mathbb{P}(X \notin \mathcal{D}_L) = 0$ where 
$$
\mathcal{D}_L = \left\{x \in \mathbb{R}^d : \mathbb{R}\times (\mathbb{R}_+^{0})^d \ni (\xi, y) \mapsto L(\xi, -\langle y, x \rangle) \mbox{ is continuously differentiable} \right\}.
$$

\item For any centered ball $B_R\subset \mathbb{R}\times  (\mathbb{R}^{0}_+)^d$ of radius $R>0$,
$$
\sup_{(\xi,y)\in B_R} \left\{|\partial_\xi L(\xi, -\langle y, X\rangle)| + |X| |\partial_x L(\xi, -\langle y, X\rangle)|\right\} \in L^{1}(\mathbb{P}).
$$

\end{itemize}
Then, the map $h$ is continuously differentiable on $\mathbb{R}\times (\mathbb{R}^{0}_+)^d$ with 
$$
\partial_{\xi} h(\xi, y) = \mathbb{E}[\partial_\xi L(\xi, -\langle y, X\rangle)] \quad \mbox{ and }\quad \partial_{y_i} h(\xi, y) =\mathbb{E}\Big[-X_i \partial_x L(\xi, -\langle y, X\rangle)-\frac{b_i}{y_i}\Big], \quad i=1, \cdots, d,
$$
\noindent so that  
$$
\Argmin h = \left\{ z\in \mathbb{R}\times (\mathbb{R}^{0}_+)^d : \nabla h(z) = 0 \right\}.
$$

Moreover, if $(\xi^\star, y^\star) \in \Argmin h$, then $y^{\star}$ is the unique minimizer of $\Gamma_g$ so that $u^\star= \frac{y^\star}{\|y^\star\|_1}$ is the unique solution to the risk budgeting problem RB$_b$. 
\end{proposition}

The above proposition suggests that, to solve the original risk budgeting problem, we must address the stochastic optimization problem \eqref{reformulation:rb:problem:bayes:risk}. To this end, we develop a stochastic Mirror Descent (SMD) algorithm with limiting point in $\left\{ \nabla h = 0 \right\}$.

As it can be noticed, somehow we end up with the same difficulty as before since the gradient of $H(., X)$ is unbounded due to the presence of the term $\sum_{i=1}^{d}b_i\log(y_i)$. We thus apply the same taming technique as in the deterministic framework, namely, similarly to what we observed in Section \ref{sub:sec:taming:singularity}, it holds
{\small
\begin{align*}
& \left\{ z \in \mathbb{R} \times (\mathbb{R}^{0}_+)^d : \nabla h(z) = \mathbb{E}\left[ \begin{pmatrix} \partial_{\xi} H(z, X) \\ \nabla_y H(z, X) \end{pmatrix}  \right] =0 \right\} \\
& \quad \quad =   \left\{ z \in \mathbb{R} \times (\mathbb{R}_+)^d : \mathbb{E}\left[ \begin{pmatrix} \partial_{\xi} H(z, X) \\ \kappa(y) \nabla_{y} H(z, X) \end{pmatrix} \right] =0 \right\}.
\end{align*}
}
\noindent recalling that $\kappa(y) = \underline{y}\wedge 1$ and $\underline{y}= \min_{1\leq i \leq d} y_i$.

Indeed, similarly to the deterministic case, if $z \in \mathbb{R} \times (\mathbb{R}_+^d \backslash (\mathbb{R}^{0}_+)^d)$ then $\underline{y}=0$ so that $\kappa(y) \partial_{y_i}H(z, X) = - b_i \neq 0$ if $y_i=0$ and $0$ otherwise, which in turn implies 
$$ 
\mathbb{E}\left[ \begin{pmatrix} \partial_{\xi}H(z, X) \\ \kappa(y) \nabla_{y}H(z, X) \end{pmatrix} \right] \neq 0 \text{ for } z \in \mathbb{R} \times (\mathbb{R}_+^d \backslash (\mathbb{R}^{0}_+)^d). 
$$

Analogously to the previous approach, we will construct our SMD algorithm using $(\partial_{\xi}H(z, X), \kappa(y) \nabla_{y}H(z, X))$ instead of $\nabla H(z, X)$. The following result, akin to Lemma \ref{lem:mean:reverting} follows from the convexity of $h$, and its proof is therefore omitted. 

\begin{lemma} \label{mean:reverting:stochastic}
For all $z \in \mathbb{R} \times \mathbb{R}_+^d, \ z\neq z^\star:=(\xi^{\star}, y^{\star})$, it holds
$$
\left\langle z-z^\star, \mathbb{E}\left[\begin{pmatrix} \partial_{\xi} H(z, X) \\ \kappa(y) \nabla_{y}H(z, X) \end{pmatrix} \right] \right\rangle > 0.
$$
\end{lemma}

\subsubsection{Stochastic Mirror Descent algorithm}\label{subsubsec:stochastic:MD}

In order to deal with the couple $z=(\xi, y)$, we will work with the so-called Bregman divergence over pairs defined by
$$
D_G(z, z')= \frac12 (\xi-\xi')^2 + D_F(y,y'), \quad z, z' \in \mathbb{R} \times (\mathbb{R}^{0}_+)^d,
$$

\noindent associated to the map
$$
G(z) = \frac12 \xi^2 + F(y), \quad z=(\xi, y) \in \mathbb{R} \times B_m,
$$

\noindent recalling that $F$ stands for the negative entropy function. From \eqref{ineq:strong:convexity}, we readily get
\begin{equation}\label{ineq:bregman:div:G}
   D_G(z, z') \geq
\frac12 (\xi - \xi')^2 + \frac{1}{2m} \|y-y'\|_1^2 \geq \frac{1}{4 (m\vee 1)} \|z-z'\|^2_1, \quad z,z' \in \mathbb{R}\times (\mathbb{R}_+^d\cap B_m). 
\end{equation}

Let $(\gamma_k)_{k\geq1}$ be a deterministic positive learning sequence as before. Starting from an initial point $z^0=(\xi^0,y^0) \in \mathbb{R} \times (\mathbb{R}^{0}_+)^d$, the SMD algorithm generates the sequence $(z^k)_{k\geq1}$ inductively. At each iteration $k$, given $z^k = (\xi^k, y^k) \in \mathbb{R} \times ((\mathbb{R}^{0}_+)^d \cap B_m)$, the update vector $z^{k+1} = (\xi^{k+1}, y^{k+1})$ is obtained by solving the following optimization problem:
$$
z^{k+1} = {\rm P}^m_{z^k}\left(\gamma_{k+1} \begin{pmatrix} \partial_{\xi}H(z^k, X^{k+1}) \\ \kappa(y^k) \nabla_{y} H(z^k, X^{k+1}) \end{pmatrix} \right)
$$

\noindent where, for a given $z \in \mathbb{R} \times ((\mathbb{R}^{0}_+)^d \cap B_m)$, the proxymal mapping $P^m_z: \mathbb{R}\times\mathbb{R}^d \rightarrow \mathbb{R} \times ((\mathbb{R}^{0}_+)^d \cap B_m )$, associated to $D_G$, is defined by
\begin{equation}\label{eq:proxymal:mapping}
{\rm P}^m_{z}(v)=\argmin_{w \in \mathbb{R} \times ( (\mathbb{R}^{0}_+)^d \cap B_m)}\left\{  \langle  v, w-z \rangle + D_G(z,w)\right\},
\end{equation}

\noindent and $(X^{k})_{k\geq 1}$ is a sequence of i.i.d. copies of $X$, independent of $z^0$.

Analogous to the deterministic case, the unique solution to the above strongly convex minimization problem is explicit. Specifically, for $k\geq0$, the update rule for $z^{k+1} = (\xi^{k+1}, y^{k+1})$ is given by: 
\begin{equation}\label{def:sequence:smd}
  \left\lbrace\begin{aligned}
&\xi^{k+1} = \xi^k - \gamma_{k+1} \partial_{\xi} H(z^{k}, X^{k+1})\\
&y^{k+1} = {\rm P}^m_{y^k}(\gamma_{k+1} \kappa(y^k) \nabla_y H(z^k, X^{k+1})) 
\end{aligned}\right.  
\end{equation}

\noindent where, for $y\in (\mathbb{R}_+^{0})^d$, the proxymal map ${\rm P}^m_{y}$ is defined as in \eqref{def:proximal:map}. We also introduce the natural filtration $\mathcal{F}=(\mathcal{F}_k)_{k\geq0}$, $\mathcal{F}_k = \sigma (\xi_0, y_0, X^1, \dots, X^k)$, $k\geq 0$, associated with the SMD scheme $(z^k)_{k\geq0}$.

The following theorem presents the central result of this section, establishing the $a.s.$ convergence of the sequence $(z^n)_{n\geq0}$, alongside an $a.s.$ convergence rate for the associated weighted averaged sequence $(\bar{z}^n)_{n\geq1}$ along its trajectory. The proof of this result is deferred to Appendix \ref{subsec:thm:conv:SMD:algorithm}. 

\begin{theorem}\label{thm:conv:SMD:algorithm} Assume that $m\geq \|y^{\star}\|_1$, that the regularity and integrability conditions on the function $L$ of Proposition~\ref{prop:rb:problem:stochastic:framework} are satisfied and that
\begin{equation}\label{growth:integrability:conditions}
\sup_{(\xi,y) \in \mathbb{R} \times (\mathbb{R}_+^d  \cap B_m)} \left\{\mathbb{E}[(\partial_\xi L(\xi,-\langle y, X\rangle))^2] +  \mathbb{E}[ X_i^2 (\partial_{x} L(\xi, -\langle y, X\rangle))^2]\right\} < \infty, \quad 1\leq i \leq d.
\end{equation}

 Assume additionally that the learning step sequence $(\gamma_n)_{n\geq1}$ satisfies $\sum_{n\geq1}\gamma_n=\infty$ and $\sum_{n\geq1} \gamma_n^2< \infty$. Then, the sequence $(z^n)_{n\geq0}$ defined by \eqref{def:sequence:smd} converges $a.s.$ to the unique minimum $z^{\star}$ of $h$ as $n\uparrow \infty$. Moreover, the weighted averaged sequence $(\bar{z}^{n})_{n\geq 1}$ of $(z^n)_{n \geq 0}$ defined by
$$
\bar{z}^n= \frac{\sum^n_{k=1} \gamma_k z^{k-1}}{\sum^n_{k=1} \gamma_k}, \quad n\geq 1,
$$
\noindent satisfies the following $a.s.$ upper-bound
\begin{equation}\label{conv:rate:stochastic:md}
h(\bar{z}^{n}) - h(z^{\star}) \leq \frac{\frac12 (\xi^0-\xi^\star)^2 + D_F(y^\star, y^0) - M_n + \frac12 m\vee 1 \sum^{n}_{k=0} \gamma_{k+1}^2 Y_{k+1}}{(\min_{0\leq k\leq n-1} \underline{y}^k \wedge 1) \sum^{n}_{k=0} \gamma_{k+1}},
\end{equation}

\noindent where $(M_n)_{n\geq1}$ is an $\mathcal{F}$-martingale satisfying $\sup_{n\geq1}\mathbb{E}[M_n^2]<\infty$ (thus converging $a.s.$ as $n\uparrow \infty$ to $M_\infty<\infty$) and $(Y_n)_{n\geq 1}$ is a non-negative sequence satisfying $\sup_{n\geq0}\mathbb{E}[Y_{n+1}|\mathcal{F}_n] \leq N^2_{\star}$ $a.s.$ for some constant $N^2_{\star}<\infty$.
\end{theorem} 

\begin{remark} Analogous to Remark \ref{remark:initialisation:algo}, selecting $y_0=e^{-1}(1, \cdots, 1)$ if $m\geq e^{-1}d$ or $y_0=m d^{-1}(1,\cdots, 1)$ otherwise, provides explicit control over the initialization error: $D_F(y^\star, y^0)\leq (m\log m)_+ + \frac{d}{e}$.
    
\end{remark}

\begin{remark} Note that standard convergence rate results for the sequence $(\bar{z}^n)_{n\geq1}$ are typically expressed as an upper bound on $\mathbb{E}[h(\bar{z}^{n}) - h(z^{\star})]$ of order $(\sum_{k=0}^{n}\gamma_{k+1})^{-1}$, under the conditions $\sum_{n} \gamma_n=\infty$ and $\sum_{n} \gamma_n^2<\infty$; see, for instance,  \cite{nemirovskij1983problem}. In our case, however, the upper-boudn \eqref{conv:rate:stochastic:md} gives
$$
\limsup_{n} \, \big(\sum_{k=0}^{n} \gamma_{k+1}\big) \times (h(\bar{z}^n)-h(z^\star))<\infty \quad a.s. 
$$

\noindent due to the fact that $\sup_{n\geq1} M_n < \infty$, $\sum_{n\geq1} \gamma_{n+1}^2 Y_{n+1}<\infty$ and $\min_{n\geq0} \underline{y}^n>0$ $a.s.$ While it would be valuable to establish an upper bound for $\mathbb{E}[h(\bar{z}^{n}) - h(z^{\star})]$, this proves to be quite challenging, as it would require an $L^{p}(\mathbb{P})$-control, for some $p>1$, on $(\min_{0\leq k\leq n} \underline{y}^k \wedge 1)^{-1}$ uniformly in $n$. 
\end{remark}

\begin{remark} Similarly to the deterministic framework, see Remark \ref{remark:selection:m}, it is essential to select $m$ sufficiently large, ensuring that $m\geq \|y^{\star}\|$. Given that $y^\star$ remains unknown, this necessitates a somewhat arbitrary or blind choice on the part of the user. While an excessively large $m$ may influence the upper bound, we find that, in practical implementations, it does not significantly impact the convergence rate of the weighted averaged sequence $(\bar{z}^n)_{n\geq1}$. We again refer the reader to Section \ref{section:numerics} for a discussion of the numerical evidence regarding the impact of the choice of $m$.
\end{remark}

\subsubsection{Some examples: Volatility, Expected Shortfall, Deviation measures}\label{subsection:examples}

The most commonly used risk measure for the risk budgeting problem is certainly the volatility. Taking $g(x)=x^2$, $L(\xi, x) = (x-\xi)^2$ and assuming that $X \in L^2(\mathbb{P})$, we observe that
\begin{equation}
\label{eq:vol}
g(r_\rho(y)) = y^T\Sigma y = \min_{\xi \in \mathbb{R}}\mathbb{E}[L(\xi,-\langle y, X\rangle)].
\end{equation}

Thus, volatility as a risk measure naturally fits within the stochastic framework outlined in Section \ref{subsubsec:stochastic:framework}. In particular, Proposition \ref{prop:rb:problem:stochastic:framework} holds, with $\Argmin h= \left\{ (\xi^\star, y^\star) \right\}$, $\xi^\star = \mathbb{E}[-\langle y^\star, X\rangle]$ and $y^\star$ is the unique minimizer of $\Gamma_g$ when volatility is employed as the risk measure. Note, however, that the condition \eqref{growth:integrability:conditions} in Theorem \ref{thm:conv:SMD:algorithm} is not met, since $\xi \mapsto \mathbb{E}[(\partial_\xi L(\xi, -\langle y, X\rangle))^2]$ and $\mathbb{E}[(\partial_x L(\xi, -\langle y, X\rangle))^2]$ both exhibit quadratic growth in $\xi$, uniformly in $y \in \mathbb{R}_+^d \cap B_m$. Nevertheless, the proof of Theorem \ref{thm:conv:SMD:algorithm} can be readily adapted to overcome this minor difficulty, ensuring $a.s.$ convergence of the sequence $(z^n)_{n\geq1}$, along with a similar convergence rate for $(\bar{z}^n)_{n\geq1}$. We shall not dwell further on the adaptation of this proof.

Our second example is the Expected Shortfall (ES), calculated at a given confidence level $\alpha \in (0,1)$. This risk measure is closely related to another well-known measure, the Value-at-Risk (VaR). Both ES and VaR are among the most widely employed risk measures in the finance and insurance industries. As in~\cite{FS10,AT02}, for a real-valued random vector $Z \in L^{1}(\mathbb{P})$, these two risk measures are defined by
\begin{equation}
\VaR_\alpha(Z)
= \inf{\big\{\xi\in\mathbb{R} : \mathbb{P}(Z\leq\xi)\geq\alpha\big\}},
\qquad
\ES_\alpha(Z)
 = \frac1{1-\alpha}\int_\alpha^1\VaR_a(Z) \, \mathrm{d}a.
\end{equation}

As stated in~\cite{RU00,BT07,BFP09:2,BFP09:1}, the VaR and ES are linked by a convex optimization problem. If the cdf $F_{Z}$ of $Z$ is continuous, then $\VaR_\alpha(Z)$ is the left-end solution to
\begin{equation}
\label{eq:sa:opt}
\argmin_{\xi\in\mathbb{R}}\left\{V(\xi):= \mathbb{E}[H(\xi, Z)] \right\},  
\quad\mbox{ where }\quad
L(\xi, Z) := \xi+\frac1{1-\alpha} (Z-\xi)^+, \quad  \xi\in\mathbb{R}.
\end{equation}
Moreover, $V$ is convex and continuously differentiable on $\mathbb{R}$, with $V'(\xi)=\mathbb{E}[\partial_\xi L(\xi, Z) ] = \frac1{1-\alpha}(F_{Z}(\xi)-\alpha)$, $\xi\in\mathbb{R}$. If $F_{Z}$ is additionally increasing, then $V$ is strictly convex and $\VaR_\alpha(Z)$ is the unique minimizer of $V$:
\begin{equation}
\label{eq:sa:opt:sol}
 \VaR_\alpha(Z)=\argmin{V}
\end{equation}
\noindent and
\begin{equation*}
\ES_\alpha(Z)=\min{V}. 
\end{equation*}

Besides, if $Z$ admits a continuous pdf $f_{Z}$, then $V$ is twice continuously differentiable on $\mathbb{R}$, with $V''(\xi)=\frac1{1-\alpha}f_{Z}(\xi)$, $\xi\in\mathbb{R}$.

Hence, assuming that the random vector $X\in L^{1}(\mathbb{P})$ and that, for any $y \in (\mathbb{R}^0_+)^d$, the cdf of $-\langle y, X\rangle\in L^{1}(\mathbb{P})$ is continuous and increasing, we see that the ES fits our stochastic framework of Section \ref{subsubsec:stochastic:framework}. In particular, Proposition \ref{prop:rb:problem:stochastic:framework} is valid with $\Argmin h = \left\{(\xi^\star, y^\star) \right\}$ where $y^\star$ is the unique minimizer of $\Gamma_g$ associated to the ES and $\xi^\star = \VaR_\alpha(-\langle y^\star, X\rangle)$. Moreover, Theorem \ref{thm:conv:SMD:algorithm} is also satisfied.

Our final example concerns the class of deviation risk measures, of which the standard deviation is undoubtedly the most widely employed instance, see, for example, \cite{rock:uryasev:zaba:06, rock:uryasev:zaba:08, ROCKAFELLAR201333}. Other examples include the mean absolute deviation and semi-deviation, each designed to capture different facets of risk. Formally, a risk measure $\rho$ belongs to this class if it satisfies \textbf{(PH)}, \textbf{(SA)}, $\rho(Z+c) = \rho(Z)$ for all random variable $Z$ and all constant $c$ and $\rho(Z)>0$ for any $Z\neq 0$. Building on \cite{cetingoz2023risk}, we consider RB-compatible risk measures that encompass both symmetric and asymmetric deviation measures, to which the stochastic framework outlined in Section \ref{subsubsec:stochastic:framework} applies. For $a, b>0$, and $Z \in L^{p}(\mathbb{P})$, for some $p\geq1$, we let
\begin{equation}\label{dev:risk:measure}
    \rho(Z) = \min_{\xi \in \mathbb{R}} \mathbb{E}[L(\xi, Z)]^{1/p}, \quad \mbox{ with } \quad L(\xi, z) := (a (Z-\xi)_+ + b (Z-\xi)_-)^p.
\end{equation}

Then, $\rho$ is an RB-compatible risk measure. The aforementioned family encompasses several well-known risk measures for specific choices of $a, \, b$ and $p$. When $a=b=1$, it yields symmetric measures such as the mean absolute deviation around the median (MAD) for $p=1$ or the standard deviation for $p=2$. In cases where $a\neq b$, we obtain asymmetric measures. For instance, if $a = \frac{\alpha}{1-\alpha}$, $b=1$ and $p=1$, the resulting measure is $\rho(Z)=\ES_\alpha(Z)-\mathbb{E}[Z]$. Similarly, with  $a=\sqrt{\alpha}$, $b=\sqrt{1-\alpha}$ and $p=2$, we derive the square-root of the variantile at level $\alpha$.

Observe that the map $\mathbb{R} \ni \xi \mapsto  \mathbb{E}[L(\xi, Z)]$ is convex. Moreover, if the cdf of $Z$ is continuous then the dominated convergence theorem guarantees that it is continuously differentiable so that
$$
\Argmin \mathbb{E}[L(., Z)] = \left\{ \xi \in \mathbb{R} : \mathbb{E}[\partial_\xi L(\xi, Z)] = 0 \right\}
$$
\noindent where
$$
\partial_\xi\mathbb{E}[L(\xi, Z)] = \mathbb{E}[\partial_\xi L(\xi, Z)]= p \mathbb{E}[-a^p (Z-\xi)_{+}^{p-1} + b^p (Z-\xi)_{-}^{p-1}] 
$$
\noindent with the convention $(Z-\xi)_+^0 = \textbf{1}_{Z\geq \xi}$ and $(Z-\xi)_-^0 = \textbf{1}_{Z\leq \xi}$. As with the example of ES, the set $\Argmin  \mathbb{E}[L(., Z)]$ does not generally reduce to a singleton. However, if the cdf of $Z$ is increasing then it does. Specifically, for $p=1$, we have $\Argmin \mathbb{E}[L(., Z)] = \left\{ \VaR_{a/(a+b)}(Z)\right\}$. 

Now, assuming that the random vector $X\in L^{p}(\mathbb{P})$ and that, for any $y \in (\mathbb{R}^0_+)^d$, the cdf of $-\langle y, X\rangle$ is continuous and increasing,  we observe that $\rho$ aligns with the stochastic framework of Section \ref{subsubsec:stochastic:framework} with the map $g(x)=x^p$. In particular, Proposition \ref{prop:rb:problem:stochastic:framework} holds with $\Argmin h = \left\{(\xi^\star, y^\star) \right\}$, where $y^\star$ is the unique minimizer of $\Gamma_g$ and $\xi^\star$ is the unique minimizer associated with the optimization problem \eqref{dev:risk:measure}, with $Z=-\langle y^\star, X\rangle$. If $p=1$, the condition \eqref{growth:integrability:conditions} of Theorem \ref{thm:conv:SMD:algorithm} is satisfied, ensuring the $a.s.$ convergence of $(z_n)_{n\geq 1}$, along with the $a.s.$ convergence rate of $(\bar{z}_n)_{n\geq1}$. However, for $p>1$, as in the first example on the volatility, \eqref{growth:integrability:conditions} is violated, since $\xi \mapsto \mathbb{E}[(\partial_\xi L(\xi, -\langle y, X\rangle ))^2]$ and $\mathbb{E}[(\partial_x L(\xi, -\langle y, X\rangle))^2]$ exhibit polynomial growth in $\xi$, uniformly in $y \in \mathbb{R}_+^d \cap B_m$. Nonetheless, one can readily adapt the proof of Theorem \ref{thm:conv:SMD:algorithm} to bypass this minor issue and establish the $a.s.$ convergence of $(z^n)_{n\geq1}$, along with a similar convergence rate for $(\bar{z}^n)_{n\geq1}$. Once again, we will refrain from delving into the technical details of this adaptation.

\section{Numerical Examples}\label{section:numerics}
\subsection{Presentation of the model and first results}\label{section:toy}
In this section, we aim to illustrate our theoretical results through a simple example.  We begin with a portfolio of three assets, constructing a risk budgeting portfolio under ES at a confidence level of $\alpha=95\%$. Equal risk budgets are assigned to each asset, i.e., $b = \left(\frac{1}{3}, \frac{1}{3}, \frac{1}{3}\right)$, which corresponds to the Equal Risk Contribution (ERC) portfolio. Our aim is to showcase the performance of our DMD and SMD algorithms by comparing the outcomes they yield against those from a conventional method.

In the analysis that follows, we assume the joint distribution of asset returns is represented by a mixture of two multivariate Student-t distributions, as this model captures asymmetric and heavy-tailed characteristics commonly observed in asset returns while allowing for precise computation of risk budgeting portfolios under ES using the semi-analytic expressions of VaR and ES \cite{cetingoz2023risk}.

Specifically, we assume that $X$ has the following density with respect to the Lebesgue measure:
$$
f_X(x) := p f(x|\mu_1, \Lambda_1, \nu_1) + (1 - p) f(x|\mu_2, \Lambda_2, \nu_2)
$$
where $p$ is the probability weight for the first distribution, $f(x|\mu, \Lambda, \nu)$ denotes the density of a multivariate Student-t distribution with mean vector $\mu$, covariance matrix $\Lambda$, degrees of freedom $\nu$ given by:
$$
f(x|\mu_i, \Lambda_i, \nu_i) := \frac{\Gamma \left[ \frac{\nu_i + d}{2} \right]}{\Gamma(\nu_i/2)\nu_i^{d/2} \pi^{d/2} \det(\Lambda_i)^{1/2}} \left\{ 1 + \frac{1}{\nu_i} (x - \mu_i)' \Lambda_i^{-1} (x - \mu_i) \right\}^{-(\nu_i + d)/2}.
$$ 

We work with a realistic model that has been calibrated using daily returns of JPMorgan Chase \& Co. (JPM), Pfizer Inc. (PFE) and Exxon Mobil Corporation (XOM) over the period August 2008–April 2022. The model parameters are estimated using the expectation-maximization algorithm. We have obtained $p=0.7$,  location vectors: $\mu_1= (0.0001, 0.0002, -0.0003)^{'}$, \  $\mu_2=(0.001, 0.0005, 0.0002)^{'} $, scale matrices: $$\Lambda_1 = \begin{pmatrix}
9 \cdot 10^{-5} & 3 \cdot 10^{-5} & 5 \cdot 10^{-5} \\
3 \cdot 10^{-5} & 9 \cdot 10^{-5} & 3 \cdot 10^{-5} \\
5 \cdot 10^{-5} & 3 \cdot 10^{-5} & 1 \cdot 10^{-4}
\end{pmatrix}, \quad
\Lambda_2 = \begin{pmatrix}
4 \cdot 10^{-4} & 1 \cdot 10^{-4} & 1 \cdot 10^{-4} \\
1 \cdot 10^{-4} & 1 \cdot 10^{-4} & 6 \cdot 10^{-5} \\
1 \cdot 10^{-4} & 6 \cdot 10^{-5} & 1 \cdot 10^{-4}
\end{pmatrix}$$ and degrees of freedom $\nu_1 = 3.4$, $\nu_2 = 2.6$. Under these parameters, we compute the risk budgeting portfolio using the L-BFGS-B algorithm using the semi-analytic expressions of VaR and ES \cite{cetingoz2023risk}. The resulting portfolio $u$ together with the associated risk contributions $(u_i \partial_{u_i} r_\rho(u))_{1\leq i\leq d}$ are computed in order to confirm that it is a solution to the risk budgeting problem. The corresponding VaR and ES are also calculated. This portfolio serves as a reliable benchmark for evaluating the convergence and accuracy of our algorithms and will hereafter be referred to as the reference portfolio. The results are presented in Table \ref{tab:risk_budgeting_portfolio}.
\begin{table}[H]
    \centering
    \begin{tabular}{c|c|c}
    \toprule
    Asset & $u^{\star}_i$ & $u^\star_i \partial_i \mathcal{R}(u^\star)$ \\
    \midrule
    1 & 0.2535 & 0.01096 \\
    2 & 0.3866 & 0.01096 \\
    3 & 0.3599 & 0.01096 \\
    \bottomrule
    \end{tabular}
    
    \vspace{0.1cm} % Espacio entre las tablas
    
    \begin{tabular}{c|c|c|c}
    \toprule
VaR & 0.0193 & ES  & 0.0329 \\
    \bottomrule
    \end{tabular}
    
    \caption{Reference portfolio: weights, risk contributions, VaR and ES.}
    \label{tab:risk_budgeting_portfolio}
\end{table}

Let us now turn to the application of the DMD and SMD schemes for computing risk budgeting portfolios. 

For the implementation of our DMD algorithm, we set $ \gamma_n = n^{-0.55}$, $ m = 100$, and perform $50{,}000$ iterations. In practice, as will be illustrated later, it suffices to select an $m$ greater than $\|y^\star\|_1$. Here, $ \|y^\star\|_1 = 30.4$. The initial point is defined as $y_i^0 = \frac{1}{d \sigma_i^2} $ for $ i = 1, \dots, d $, where $ \sigma_i^2 $ denotes the variance of asset $ i $ under the first t-Student distribution, estimated using few samples of $X$. 

To address the stochastic optimization problem outlined in Section \ref{subsubsec:stochastic:framework} for ES using our SMD algorithm, we generate $ 10^6$ samples of our multivariate Student-t mixture $X$. The algorithm is run with 10 epochs and a stepsize of $\gamma_n = n^{-0.75}$ , with $m$ as in the DMD scheme. We initialize $\xi_0 = 0$ and $y^0$ as before. 

Our results are presented in Table \ref{tab:DMD_SMD_results}. We observe that the DMD algorithm accurately computes the optimal portfolio weights without error. The estimated VaR and ES values are not included, as they correspond to those in Table \ref{tab:risk_budgeting_portfolio}. Additionally, we note that the estimated weights $ u_i^{\text{SMD}}$ and the estimated VaR from the SMD algorithm closely align with those of the reference portfolio.\footnotetext[1]{While implementing a mini-batch version would likely enhance our algorithm’s accuracy, we have opted to adhere to our theoretical framework.}

\begin{table}[H]
    \centering
    \begin{tabular}{c|c|c|c}
    \toprule
    Asset & $ u_i^{\text{SMD}}$ & $ \left| u^\star_i - u_i^{\text{SMD}} \right|/u^\star_i$ & $ u_i^{\text{DMD}}$ \\
    \midrule
    1 & 0.2537 & 0.08\% & 0.2535 \\
    2 & 0.3877 & 0.30\% & 0.3866 \\
    3 & 0.3586 & 0.40\% & 0.3599 \\
    \bottomrule
    \end{tabular}
    
    \vspace{0.1cm} % Espacio entre las tablas
    
    \begin{tabular}{c|c|c|c}
    \toprule
    VaR$^{\text{SMD}}$ & 0.0194 & (VaR$^{\text{SMD}}$ - VaR)/ VaR & 0.52\% \\
    \bottomrule
    \end{tabular}
    
    \caption{Estimated weights and VaR together with relative difference using SMD algorithms and estimated weights using the DMD algorithm.}
    \label{tab:DMD_SMD_results}
\end{table}

\begin{figure}[h] 
    \centering
    \includegraphics[width=1 \textwidth]{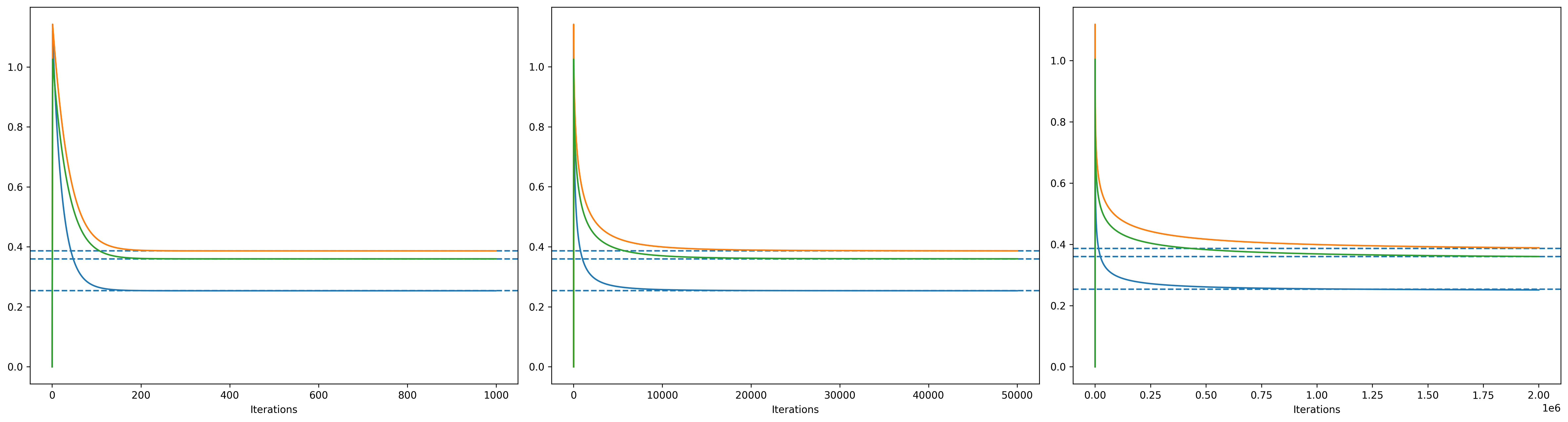}
    \caption{Evolution of the three components of $(u^k)_{k\geq0}$ of the DMD algorithm for different step sequence $(\gamma_n)_{n\geq1}$. Left: $\gamma_n \equiv 1$. Center: $\gamma_n = n^{-0.55}$. Right: $\gamma_n = n^{-0.75}$. Dashed lines are the asset weights of the reference portfolio in all subplots.}
    \label{ysdet}
\end{figure}

We now observe the convergence by plotting the three components of the sequence $(u^k)_{k \geq 0}$ generated by the DMD algorithm in Figure \ref{ysdet} with different step sizes: $\gamma_n = 1$, $\gamma_n = n^{-0.55}$, $\gamma_n=n^{-0.75}$, $n\geq1$.

It is worth noting that the choice of step sequence $ (\gamma_n)_{n \geq 1} $ significantly impacts the speed of convergence. Setting $\gamma_n = \gamma = 1 $, i.e., a fixed step size of 1 across iterations, appears to be optimal compared to reducing the step size over time, even though our theoretical results do not guarantee convergence in this specific instance. Indeed, convergence is reached in fewer than 1,000 iterations with $\gamma_n = 1 $, whereas choosing $ \gamma_n = n^{-0.55} $ ensures convergence within 50,000 iterations. By contrast, selecting \( \gamma_n = n^{-0.75} \) represents the worst case. This is in line with the upper-bound provided in Theorem \ref{thm:conv:DMD}. These results highlight the significant impact of the selection of the step sequence $(\gamma_n)_{n\geq1}$ on the convergence rate of the DMD algorithm.

\begin{figure}[h] 
    \centering
    \includegraphics[width=1 \textwidth]{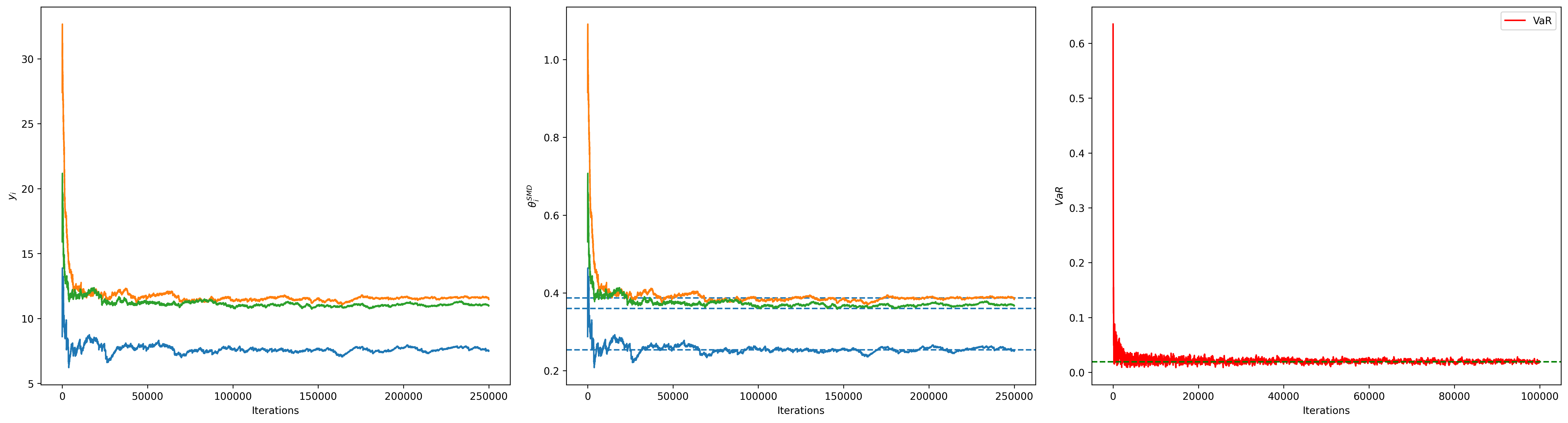}
    \caption{Convergence of the SMD algorithm. Left: evolution of the three components of $(y^k)_{0\leq k\leq N}$. Center: evolution
of the three components of $(u^{k})_{0\leq k \leq N}$ — dashed lines are the weights of the reference portfolio.
Right: evolution
of $(\xi^k)_{0\leq k \leq N'}$, with $N'=100,000$ — dashed line is the VaR of the reference portfolio }
    \label{ys:smd:gamma:low}
\end{figure}

We now run the SMD algorithm with $N = 250,000$ iterations, comprising 25,000 samples and 10 epochs. We initialize $(y^0, \xi^0)$ and set $m$ as before, with $\gamma_n = n^{-0.55}$ for $ n \geq 1 $. Convergence is observed by plotting the evolution of the components of both sequences $(y^k)_{0 \leq k \leq N}$ and  $(u^k := y^k / \|y^k\|_1)_{0 \leq k \leq N} $, alongside $ (\xi^k)_{0 \leq k \leq N} $ in Figure \ref{ys:smd:gamma:low}. A rapid convergence to the weights and VaR of the reference portfolio is achieved, with the convergence of $(\xi^k)_{k \geq 0}$ appearing faster than that of $(u^k)_{k \geq 0}$.

\begin{figure}[h] 
    \centering
    \includegraphics[width=1 \textwidth]{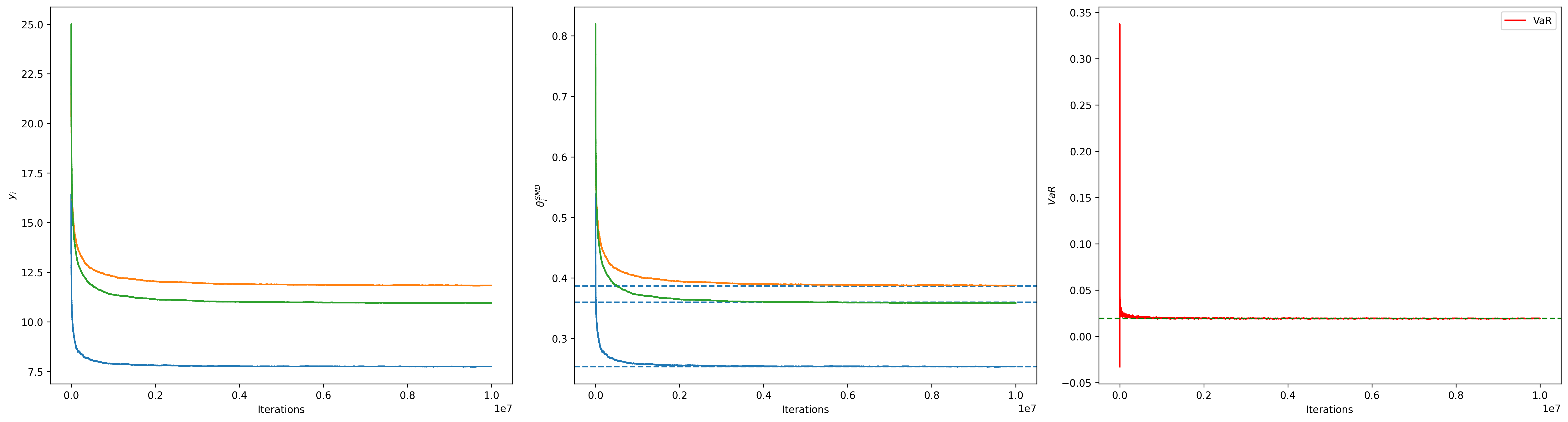}
    \caption{Convergence of the SMD algorithm. Left: evolution of the three components of $(y^k)_{k\geq0}$. Center: evolution
of the three components of $(u^k)_{k\geq0}$'s — dashed lines are the asset weights of the reference portfolio.
Right: evolution
of $(\xi^k)_{k\geq0}$ — dashed line is the VaR of the reference portfolio. }
    \label{ys:smd:gamma:high}
\end{figure}

In Figure \ref{ys:smd:gamma:high}, we increase the number of iterations in the SMD algorithm, keeping the parameters unchanged except for setting $ N = 10^7 $ iterations, divided into $ 10^6 $ samples and 10 epochs.

\begin{figure}[h] 
    \centering
    \includegraphics[width=1 \textwidth]{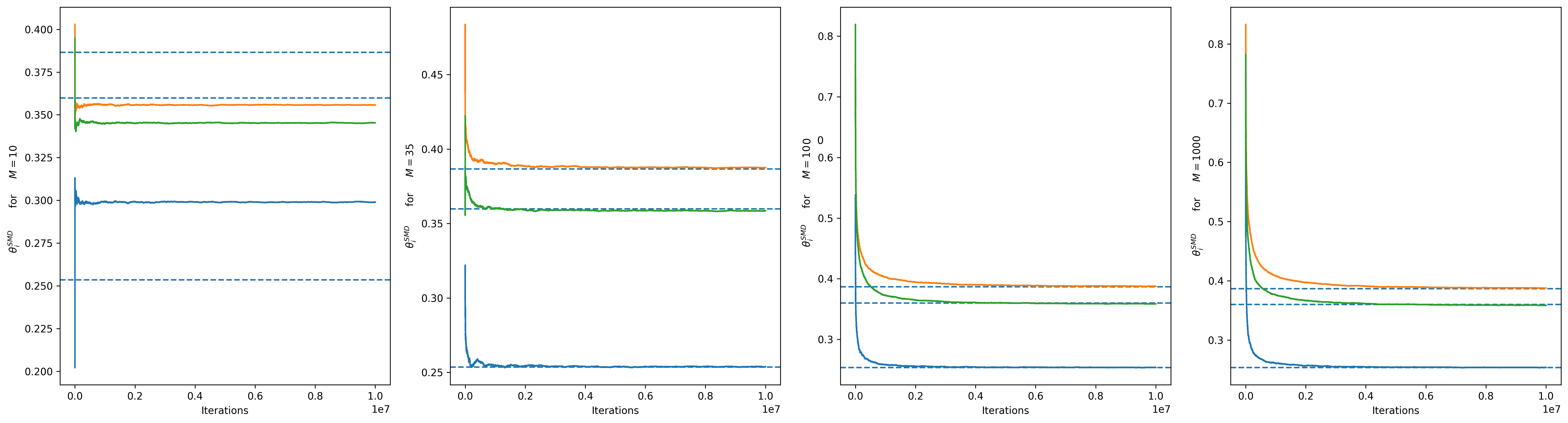}
    \caption{ From left to right: evolution of the sequence $(u^{k})_{k\geq0}$ for $m=10$, 35, 100 and 1000 respectively.}
    \label{ys:M}
\end{figure}

Note that the choice of $m$ is not critical, provided it is selected to be larger than $\|y^\star\|_1$. Consequently, we consistently use a large $m$, as we did before, since selecting an $m$ smaller than $\|y^\star\|_1$ prevents convergence. This effect is illustrated in Figure \ref{ys:M}. When $m = 10 $, we do not observe convergence of $(u^k)_{k \geq 0}$ towards $u^\star$; however, for values of $m$ larger than $\|y^\star\|_1$, convergence occurs, particularly for $ m = 35 $, 100, and 1000, with the results for $m = 100$ and $m = 1000 $ appearing almost identical due to minimal projections during the algorithm. The DMD algorithm exhibits similar behavior.

We proceed with our analysis by assessing the robustness of both MD algorithms with respect to the dimension $d$, corresponding to the number of assets. We consider four scenarios, calibrating our model with 10, 50, 100, and 250 assets (see Section~\ref{taming_subsec} for details on selecting an arbitrary number of assets $d$ and the calibration process). To demonstrate the convergence of both schemes, we calculate the mean deviation error (MDE) in asset weights compared to those of the reference portfolio, defined as
$$
\textnormal{MDE} = \frac{1}{d}\sum_{i=1}^d |u^{N}_i-u_i^{\star}|. 
$$

 For the DMD algorithm, we use a step sequence $\gamma_n = n^{-0.55}$ and $50,000$ iterations. For the SMD algorithm, we set the parameters as follows: $\gamma_n = n^{-0.75}$, with $10^6$ samples combined with 10 epochs. The parameters $\xi^0, \ y^0 $ and $m$ remain as previously specified. The results are presented in Table \ref{tab:multiassets:SMD:results}. Note that values for SMD and DMD are scaled by a factor of $10^{3}$ for clarity. The relative error of VaR is presented as a percentage.

\begin{table}[h]
    \centering
    \begin{tabular}{c|c|c|c}
    \toprule
    Assets & \multicolumn{1}{c|}{SMD} & \multicolumn{1}{c|}{$\left| \text{VaR}^{\text{SMD}} - \text{VaR} \right| / \text{VaR}$} & \multicolumn{1}{c}{DMD} \\
    \midrule
    10  & 1.02 & 0.005\% & 0.19 \\
    50  & 0.26 & 0.980\% & 0.76 \\
    100 & 0.25 & 0.075\% & 1.15 \\
    250 & 0.26 & 0.977\% & 0.85 \\
    \bottomrule
    \end{tabular}
    \caption{MDE for both DMD and SMD algorithms, scaled by $10^{3}$, along with the relative error of VaR with respect to the reference portfolio.}
    \label{tab:multiassets:SMD:results}
\end{table}

We also performed computations for the DMD algorithm using a constant step size of $\gamma_n=1$ over 10,000 iterations. These results are not included in Table \ref{tab:multiassets:SMD:results}, as no errors were observed in any instance, underscoring the strong robustness of the DMD algorithm when a constant step size is applied. The findings for both SMD and DMD with a decreasing step size similarly demonstrate impressive robustness, as they maintain convergence even with increasing dimensionality. For the variable step size case, the performance remains generally satisfactory, though slightly lower than that achieved with a fixed step size, as convergence is still ongoing and has not yet fully stabilized.

\subsection{Error analysis for different portfolio sizes}
\subsubsection{Taming to the rescue: fixing stability issues}
\label{taming_subsec}
In the traditional approach to solving the risk budgeting problem, the numerical algorithm typically follows the directions defined by the gradient $\nabla \Gamma(y)$, recalling that $g=\text{Id}$ for the case of ES. However, since the gradient diverges on $\partial (\mathbb{R}_+)^d$, as proposed in Section \ref{sub:sec:taming:singularity}, we moderate it by applying the factor $\kappa(y)$ to ensure convergence of the algorithm. In this subsection, we aim to demonstrate that, beyond its theoretical benefits, this adjustment also greatly enhances the stability and convergence of the numerical scheme, particularly in stochastic settings.

We begin by comparing the SMD algorithm, following the update rule given in \eqref{def:sequence:smd} with $m=100$, to two alternative methods in terms of successful convergence across different scenarios. The first method, referred to as classical-SGD (c-SGD), is the projected stochastic gradient descent algorithm proposed in \cite{cetingoz2023risk} to solve (\ref{reformulation:rb:problem:bayes:risk}) using the update rule  
\begin{equation*}\label{def:sequence:sgd}
  \left\lbrace\begin{aligned}
&\xi^{k+1} = \xi^k - \gamma_{k+1} \partial_{\xi} H(z^{k}, X^{k+1})\\
&y^{k+1} =  \Pi\left(y^k  - \gamma_{k+1} \nabla_y H(z^k, X^{k+1})\right)
\end{aligned}\right.
\end{equation*}

 \noindent recalling that $z^k = (\xi^k, y^k)$ and $\Pi : \mathbb{R}^d \rightarrow (\mathbb{R}^{0}_+)^{d}$ is a projection function designed to ensure that all elements are positive. Specifically, $\Pi$ is defined to replace any negative elements with a fixed positive value, $\epsilon =  10^{-4}$.

The second approach, referred to as tamed-SGD (t-SGD), incorporates the concept of taming the gradient and involves updating $(z^k)_{k\geq0}$ as follows:
\begin{equation*}\label{def:sequence:sgd_tamed}
  \left\lbrace\begin{aligned}
&\xi^{k+1} = \xi^k - \gamma_{k+1} \partial_{\xi} H(z^{k}, X^{k+1})\\
&y^{k+1} =   \Pi\left(y^k  - \gamma_{k+1} \kappa(y^k) \nabla_y H(z^k, X^{k+1})\right)
\end{aligned}\right. 
\end{equation*}

\noindent recalling that $\kappa(y)=\underline{y}\wedge 1$ serves as the taming factor, applied in both DMD and SMD algorithms to adjust the gradient and enhance stability.

At this stage, our primary focus is on assessing the algorithms’ ability to approach the true solution across various portfolio sizes, rather than precisely measuring the accuracy of their convergence. This preliminary evaluation offers insight into the algorithms' convergence behavior, setting the groundwork for a subsequent analysis of their accuracy, which will be addressed in the following subsection. In brief, the objective here is to understand the practical significance of gradient taming in mitigating numerical divergence caused by the issue of exploding gradients.

We proceed with computing ERC portfolios for ES. To statistically analyze the convergence rate of different algorithms across various portfolio sizes $d$, we follow this procedure. We randomly select $d$ stocks from the S\&P 500 components (as of Q2 2022) and obtain their daily returns from August 2008 to April 2022. To capture the distributional characteristics of these returns, we fit a two-component Student-t mixture model to the historical data. Based on the estimated parameters, we compute a reference portfolio. We then simulate $10^6$ data points from the fitted model and apply the SMD, c-SGD, and t-SGD algorithms to the simulated dataset, analyzing convergence by comparing the results to the reference portfolio.

This procedure is repeated 100 times for each $d$, with a new random selection of $d$ assets in each iteration. Across these repetitions, we monitor each algorithm's rate of successful convergence, noting instances of divergence due to issues such as exploding gradients. This approach provides insights into the robustness of each algorithm in realistic portfolio construction settings. Specifically, divergence is defined as occurring when a discrepancy in the objective function value is observed at the final iteration $n=10^6$. That is, if $\Gamma({y}^n)-\Gamma(y^\star)$ exceeds a prescribed threshold $\varepsilon$, the algorithm is classified as having diverged in that particular repetition.\footnote{The error is measured by $\Gamma({y}^n) - \Gamma(y^\star)$ -- the distance of the objective function value at the last iterate to the minimum -- rather than the averaged iterates $\Gamma({\bar{y}}^n) - \Gamma(y^\star)$, as discussed in Theorem~\ref{conv:rate:deterministic:md}, to mitigate the influence of large values in earlier iterations and to assess convergence within reasonable computational times.}

\begin{table}[h]
    \centering
\begin{tabular}{c|cccccc}
\toprule
{} & \multicolumn{5}{c}{$d$} \\
$\varepsilon$ & 10  & 25  & 50  & 100 & 250 \\
\midrule
$5\times10^{-2}$  &  24 &  11 &  24 &  47 &  35 \\
$5\times10^{-1}$  &  23 &  11 &  23 &  43 &  11 \\
$5\times10^{0}$  &  23 &  10 &  13 &   8 &   0 \\
$5\times10^{1}$ &  22 &   2 &   0 &   0 &   0 \\
\bottomrule
\end{tabular}
\caption{Number of divergences out of 100 repetitions for the c-SGD algorithm.}
\label{convergence_issues_sgd_table}
\end{table}

\begin{figure}[h] 
    \centering
    \includegraphics[width=1 \textwidth]{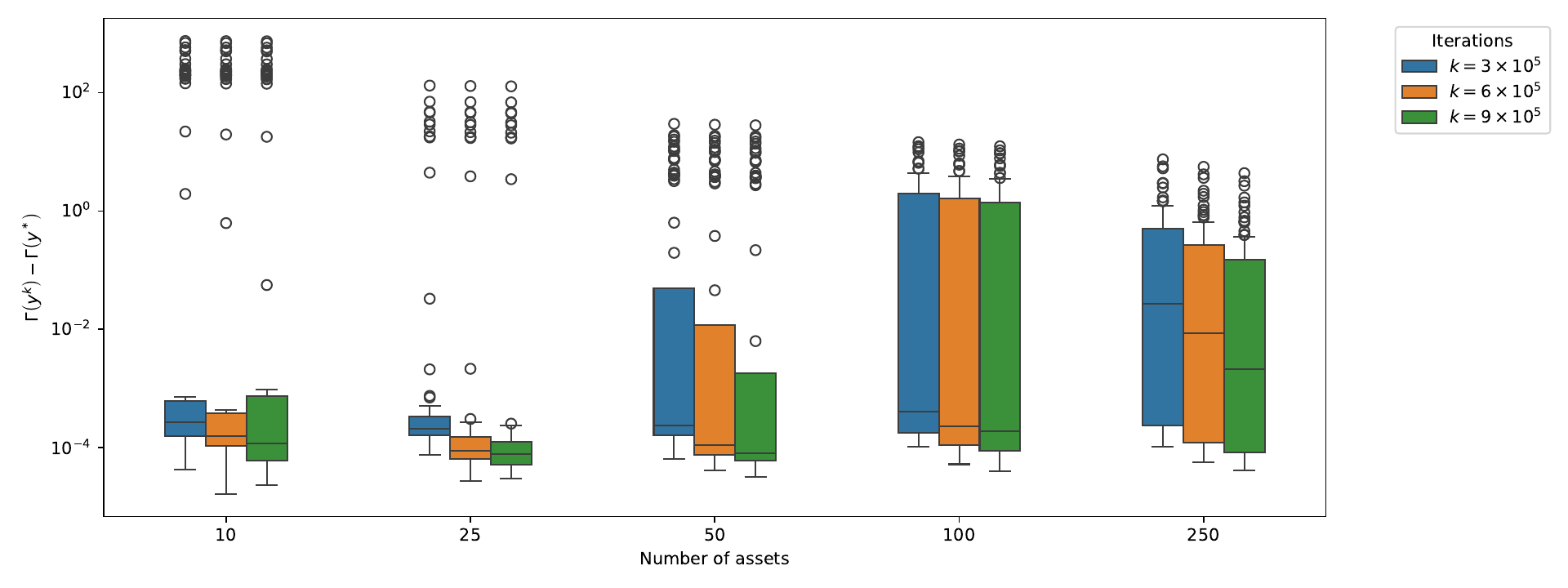}
    \caption{Errors in the objective function values for the unnormalized portfolio weights obtained via c-SGD, calculated across 100 samples for each portfolio size. The error $\Gamma({y}^k) - \Gamma(y^*)$ is measured at various iterations $k=3\times10^5, \, 6\times10^5,$ and $ 9\times10^5$ to illustrate the progression of error throughout the algorithm's execution.}
    \label{convergence_issues_sgd}
\end{figure}

Table~\ref{convergence_issues_sgd_table} underscores the inherent stability challenges associated with c-SGD and illustrates the advantages of taming the gradient.\footnote{Analogous results are not shown for t-SGD and SMD, as neither algorithm exhibited divergence across any of the cases tested.} While both t-SGD and SMD reliably converge to the reference portfolio (with precision considerations to be discussed in the following section), c-SGD encounters an issue with gradient instability due to certain elements of the vector $y$ approaching zero during iterations. This instability causes divergence in several trials. Figure~\ref{convergence_issues_sgd} further illustrates the distribution of error over 100 trials, highlighting outliers that correspond to instances of non-convergence across varying numbers of iterations $k$. The figure suggests that, although c-SGD can occasionally achieve accurate results, its convergence is highly sensitive to the initial choice of the learning rate $\gamma$.\footnote{For all methods, we employ a learning rate scheme defined as $\gamma_n =  \gamma n^{-0.65}$. For the tamed methods (t-SGD and SMD), initial learning rates $\gamma$ are set to 1, 2.5, 5, 10, and 25 for portfolio dimensions $d$ of 10, 25, 50, 100, and 250, respectively. For c-SGD, the corresponding values are set to 5, 1, 0.5, 0.25, and 0.1 for $d$ values of 10, 25, 50, 100, and 250. These learning rates have been selected through iterative tuning to optimize convergence for each algorithm and portfolio size.} In contrast, the tamed methods exhibit notable robustness to hyperparameter selection, consistently avoiding divergence across all tested cases.

\subsubsection{Improved accuracy via SMD}\label{sec:smd_t_sgd}
We now turn our attention to the SMD and t-SGD algorithms, both of which exhibit stable convergence properties in contrast to c-SGD. Our aim is to evaluate how \textit{closely} the portfolios generated by these methods approximate the reference portfolios when applied to the same dataset. This approach allows for an effective comparison of the efficiency of the update rules associated with t-SGD and SMD, while maintaining all other hyperparameters constant. By isolating the effects of these update mechanisms, we seek to determine whether SMD offers practical advantages in computing risk budgeting portfolios, in addition to its theoretical benefits.

\begin{figure}[h] 
    \centering
    \includegraphics[width=.5\textwidth]{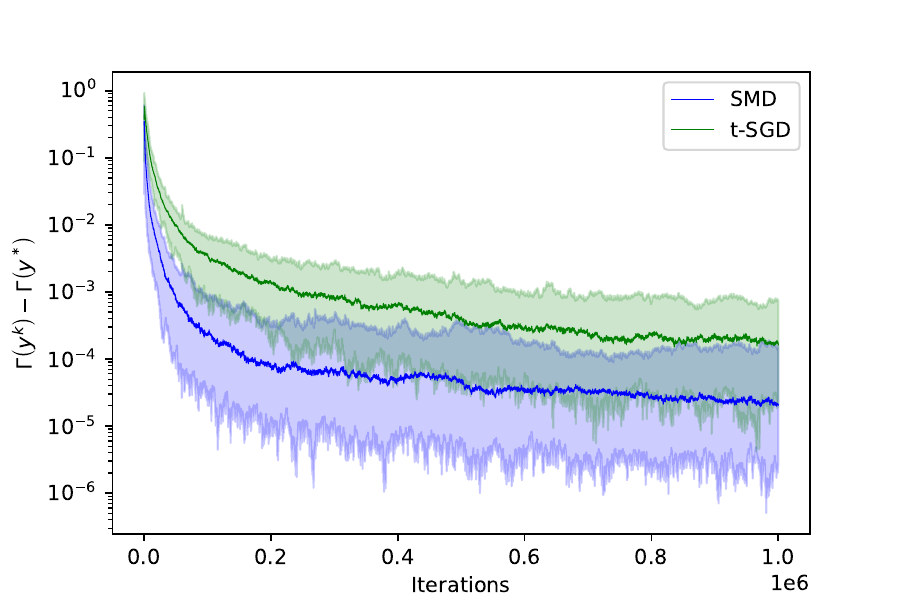}
    \caption{The evolution of error for the SMD and t-SGD algorithms across iterations.}
    \label{error_loss_iterations}
\end{figure}

An initial and illustrative assessment can be conducted by tracking the evolution of error throughout the iterations. Specifically, we can simulate data from a fixed model and run the SMD and t-SGD algorithms, recording the error $\Gamma(y^k) - \Gamma(y^*)$ after each iteration. Figure~\ref{error_loss_iterations} shows the median error along with a 95\% confidence interval (computed by running the algorithms on 100 different samples from the same model) for the 3-asset model described in Section~\ref{section:toy}. The results indicate that the SMD algorithm converges to the minimum faster than t-SGD. However, this analysis may be specific to the chosen model and should be extended to general cases and larger portfolio dimensions for a broader conclusion. 

For this reason, we apply the same procedure described in Section~\ref{taming_subsec} to assess the accuracy of the computed portfolios. Table~\ref{error_analysis_table} displays the portfolios’ accuracy relative to the reference portfolios, where accuracy is evaluated by the difference between the \textit{true} minimum objective function value and the objective function value at the computed $y^k$. Specifically, the error measure is defined as $(\Gamma(y^k) - \Gamma(y^\star)) \times 10^3$. We report the median error and median absolute deviations (in parentheses), computed by repeating the process 100 times with samples drawn from 100 distinct fitted models for each $d$. This metric serves as a quantitative measure of how closely the computed portfolios approximate the optimal solution.

\begin{table}[h]
    \centering
    \begin{tabular}{c|c|c|c|c|c|c}
    \toprule
    {} & \multicolumn{2}{c|}{$k=3\times10^5$} & \multicolumn{2}{c|}{$k=6\times10^5$} & \multicolumn{2}{c}{$k=9\times10^5$} \\
    $d$ & \multicolumn{1}{c}{SMD} & \multicolumn{1}{c|}{t-SGD} & \multicolumn{1}{c}{SMD} & \multicolumn{1}{c|}{t-SGD} & \multicolumn{1}{c}{SMD} & \multicolumn{1}{c}{t-SGD} \\
    \midrule
    10  &  0.10 (0.05) & 0.22 (0.10) & 0.06 (0.02) & 0.09 (0.04) & 0.05 (0.02) & 0.06 (0.03) \\
    25  &  0.19 (0.06) & 0.87 (0.38) & 0.11 (0.04) & 0.29 (0.15) & 0.09 (0.03) & 0.17 (0.09) \\
    50  &  0.24 (0.08) & 10.49 (2.31) & 0.13 (0.03) & 6.13 (1.65) & 0.11 (0.03) & 4.16 (1.30) \\
    100 &  0.36 (0.12) & 29.52 (3.73) & 0.16 (0.05) & 23.21 (3.21) & 0.11 (0.03) & 19.81 (2.95) \\
    250 &  0.90 (0.27) & 33.16 (9.01) & 0.43 (0.11) & 29.29 (7.93) & 0.25 (0.08) & 26.73 (7.26) \\
    \bottomrule
    \end{tabular}
    \caption{Error in objective function values obtained by SMD and t-SGD across different asset sizes, computed from samples of size $10^6$ drawn from fitted models at various iteration steps $k$.}
    \label{error_analysis_table}
\end{table}

\begin{figure}[h] 
    \centering
    \includegraphics[width=1\textwidth]{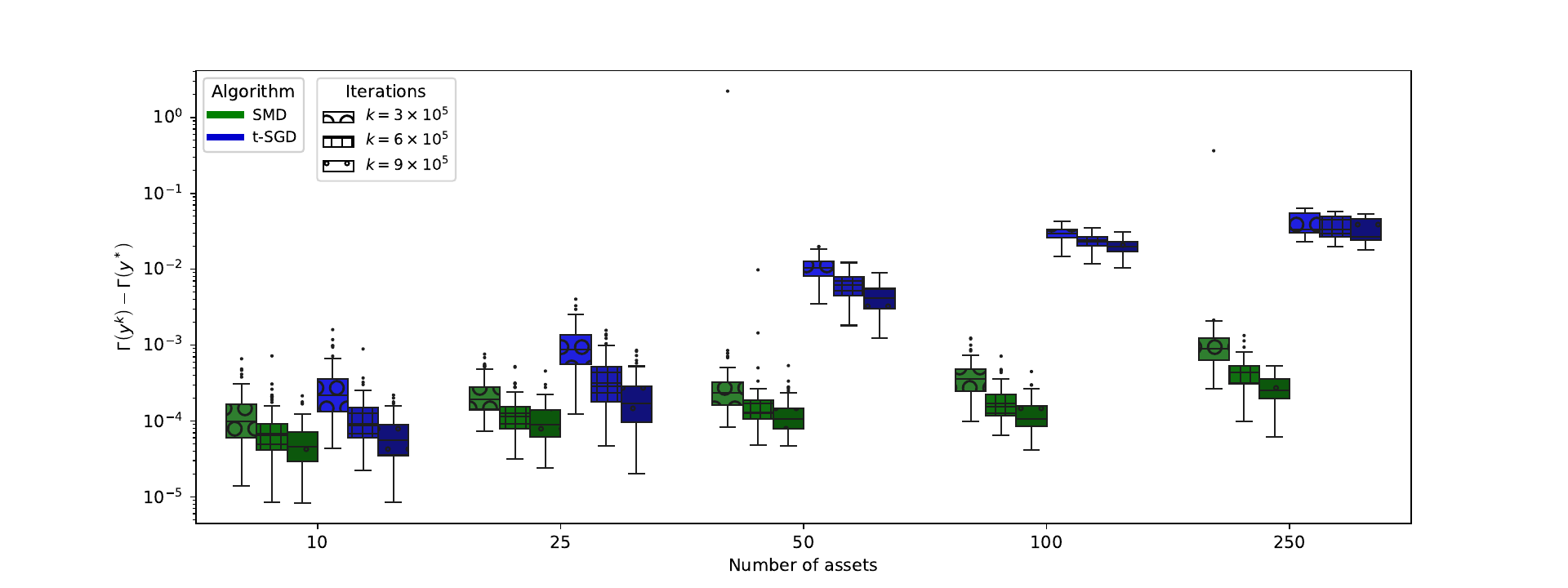}
    \caption{Distribution of errors in the objective function value obtained by using the SMD and t-SGD algorithms, computed over 100 samples for each portfolio size. }
    \label{error_loss_ds}
\end{figure}

Figure~\ref{error_loss_ds} offers further insights into the algorithms' performance. The distribution of the error in the objective function value, $\Gamma(y^k)-\Gamma(y^\star)$, is shown for the SMD and t-SGD algorithms, computed over 100 samples across various portfolio sizes: $d=10$, $25$, $50$, $100$ and $250$. The error is calculated for the weights at several iterations $k=3\times 10^5$, $6\times 10^5$, and $9\times 10^5$ to illustrate the evolution of error over the course of the algorithms. On average, the SMD algorithm consistently yields more accurate results across tested portfolio sizes and at different stages of iteration. Notably, the advantage of SMD over t-SGD becomes increasingly pronounced as portfolio size grows. A particularly interesting observation is SMD’s ability to recover from large errors over the course of iterations, with this robustness clearly seen through outliers, especially for $d=50$ and $d=250$. For instance, the outlier at iteration $k=3\times10^5$ is gradually corrected, resulting in more accurate portfolio estimations as the algorithm processes additional data points.

Although assessing error and accuracy through differences in objective function values is mathematically sound, this approach lacks practical interpretability in portfolio construction. Therefore, we extend our analysis by evaluating accuracy using the MDE ($\times 10^4$) between the computed portfolio weights and the reference portfolio. In Table \ref{error_analysis_portfolio_table}, we present the median of the MDEs alongside the median absolute deviations of the MDEs (in parentheses), calculated over 100 repetitions with samples drawn from 100 distinct fitted models for each $d$. This metric offers a more intuitive measure of practical accuracy, as it directly reflects how closely the computed portfolio weights approximate those of the reference portfolio. 

% \begin{center}
% \begin{tabular}{c|cc|cc}
% \toprule
% {} & \multicolumn{2}{c|}{Accuracy} & \multicolumn{2}{c}{Time} \\
% $d$ &          SGD &        OSBGD &          SGD &        OSBGD \\
% \midrule
% 10  &  5.46 (1.63) &  5.47 (1.65) &  1.08 (0.01) &  0.05 (0.02) \\
% 20  &  6.63 (1.82) &  6.63 (1.78) &  1.18 (0.01) &  0.14 (0.09) \\
% 50  &  7.26 (1.64) &  7.28 (1.68) &  1.32 (0.01) &  0.34 (0.12) \\
% 100 &  7.67 (1.06) &  7.69 (1.05) &  1.52 (0.01) &  0.76 (0.30) \\
% 200 &  7.60 (1.42) &  7.60 (1.41) &  1.91 (0.02) &  1.32 (0.43) \\
% 350 &  7.83 (1.73) &  7.73 (1.62) &  2.52 (0.01) &  2.53 (1.09) \\
% \bottomrule
% \end{tabular}
%  \captionof{table}{Accuracy of the Risk Budgeting portfolios obtained by the SGD and OSBGD methods for different numbers of assets under historical samples and computation time of algorithms (in seconds). The accuracy measure corresponds to $100 \lVert \theta - \theta^{method} \rVert_1$. Figures correspond to means and standard deviations (in parentheses) computed by repeating the process $m=50$ times with $\mathcal{X}_{hist}$ drawn from $m$ different $\text{DGP}_{\text{true}}$ for each $d$.}
% \label{practical_both}
% \end{center}
\begin{table}[h]
    \centering
    \begin{tabular}{c|c|c|c|c|c|c}
    \toprule
    {} & \multicolumn{2}{c|}{$k=3\times10^5$} & \multicolumn{2}{c|}{$k=6\times10^5$} & \multicolumn{2}{c}{$k=9\times10^5$} \\
    $d$ & \multicolumn{1}{c}{SMD} & \multicolumn{1}{c|}{t-SGD} & \multicolumn{1}{c}{SMD} & \multicolumn{1}{c|}{t-SGD} & \multicolumn{1}{c}{SMD} & \multicolumn{1}{c}{t-SGD} \\
    \midrule
    10  & 7.98 (1.74) & 12.09 (3.36) & 6.55 (1.32) & 8.00 (1.76) & 5.43 (1.10) & 6.06 (1.41) \\
    25  & 4.97 (0.69) & 9.52 (2.27) & 3.84 (0.50) & 5.71 (1.61) & 3.43 (0.49) & 4.74 (1.38) \\
    50  & 2.57 (0.27) & 10.73 (1.03) & 2.11 (0.22) & 8.97 (0.98) & 1.79 (0.18) & 7.94 (0.93) \\
    100 & 1.34 (0.13) & 9.19 (0.75) & 1.06 (0.11) & 8.33 (0.60) & 0.95 (0.10) & 7.81 (0.54) \\
    250 & 0.60 (0.08) & 4.91 (0.12) & 0.47 (0.05) & 4.66 (0.12) & 0.40 (0.05) & 4.49 (0.12) \\
    \bottomrule
    \end{tabular}
    \caption{MDEs of the SMD and t-SGD algorithms across different portfolio sizes, computed at iterations $k= 3\times 10^5$, $6\times 10^5$ and $9\times 10^5$.}
    \label{error_analysis_portfolio_table}
\end{table}

\begin{figure}[h] 
    \centering
    \includegraphics[width=1 \textwidth]{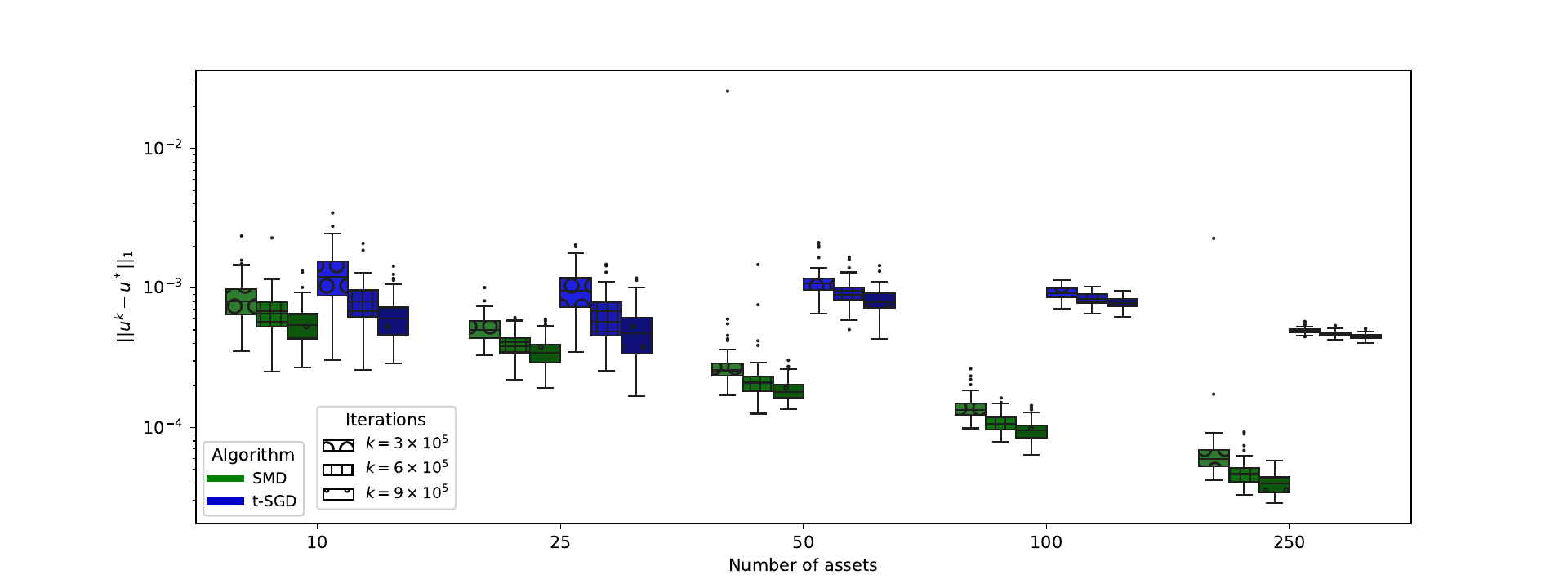}
    \caption{Distribution of the MDEs obtained by the SMD and t-SGD algorithms, computed over 100 samples for each portfolio size and iteration $n$.}
    \label{error_portfolio_ds}
\end{figure}

In Figure~\ref{error_portfolio_ds}, we display the distribution of the MDEs obtained by the SMD and t-SGD algorithms over 100 samples for each portfolio size. The error is calculated at different iterations: $k=3 \times 10^5$, $k=6 \times 10^5$ and, $k=9\times 10^5$.

Table~\ref{error_analysis_portfolio_table} and Figure~\ref{error_portfolio_ds} further confirm that the SMD algorithm produces more accurate portfolios across all tested portfolio sizes. Additionally, it achieves high accuracy even for portfolios with a large number of assets, highlighting its effectiveness as an optimization method for computing risk budgeting portfolios for ES.

\subsection{Convergence using different risk measures}
Previous numerical analyses have focused on a specific risk measure -- ES -- to have a comprehensive understanding of the behavior of SMD and SGD-based algorithms across different portfolio sizes when computing risk budgeting portfolios. However, as detailed in Section~\ref{subsection:examples}, this framework is compatible with a wide variety of risk measures. In this subsection, we aim to understand whether the computational advantage of SMD over t-SGD extends to the broader class of risk measures.

Before proceeding with the comparison of SMD and t-SGD methods using alternative risk measures within the class of deviation measures as formalized by (\ref{dev:risk:measure}) -- specifically mean absolute deviation (MAD) ($p=1$, $a=1$ and $b=1$), volatility ($p=2$, $a=1$ and $b=1$), and variantile ($p=2$, $a=0.75$ and $b=0.25$) -- it is essential to verify that the algorithms converge when these measures are used. However, we must avoid modeling $X$ with a Student-t mixture in this context, as it requires a (semi-)analytical representation of the objective function for each risk measure to compute reference portfolios accurately. Therefore, we begin with a model where $X$ follows a centered normal distribution, using the same three assets as in Section~\ref{section:toy} with the sample covariance matrix of the historical asset returns.

With this particular model, we know how to compute the risk budgeting portfolio for volatility (as defined in Equation~(\ref{eq:vol})) using deterministic algorithms such as DMD, which provides highly precise results and serves as the reference portfolio. Additionally, under the assumption of a centered normal distribution, the aforementioned risk measures should yield the same risk budgeting portfolio, as each can be represented as a function of portfolio volatility. Therefore, we simulate $10^6$ data points from this distribution and run SMD algorithm to confirm the algorithm converges to the same reference portfolio.

\begin{table}[H]
    \centering
    \begin{tabular}{c|c|c|c|c}
    \toprule
    Asset & Reference & MAD & Volatility & Variantile \\
    \midrule
    1 & 0.2411 & 0.2417 & 0.2417 & 0.2417 \\
    2 & 0.4150 & 0.4137 & 0.4138 & 0.4140 \\
    3 & 0.3438 & 0.3446 & 0.3445 & 0.3443 \\
    \bottomrule
    \end{tabular}
    \caption{Reference portfolio weights and computed weights under different portfolio measures.}
    \label{tab:rbp:risk:measures}
\end{table}

Table~\ref{tab:rbp:risk:measures} confirms that the update rule we implemented for deviation measures successfully converges to the optimal portfolio in a simple case of three assets. With this verification, we can proceed with our analysis to compare SMD and t-SGD. To make the analysis more realistic, we now model $X$ as a centered multivariate Student-t distribution with fixed degrees of freedom $\nu=4$, using the maximum-likelihood estimate for the scale matrix $\Lambda$. This model has two main advantages: the reference portfolios remain precisely computable, and the stochastic algorithms are exposed to extreme losses, making the case more representative of real-world scenarios.

In line with the procedure outlined in Section~\ref{sec:smd_t_sgd}, we assess the accuracy of the algorithms across different portfolio sizes. Specifically, we select $d$ assets from our dataset and estimate the corresponding scale matrix $\Lambda $. We then compute the reference portfolio -- identical across all risk measures -- using the centered multivariate Student-t model with the estimated $\Lambda$. From this model, we draw $n = 5 \times 10^5$ data points and apply the SMD and t-SGD algorithms. To improve the stability of the approximated portfolios, we employ the Polyak-Ruppert averaging technique. Rather than using the final iterate $y^n$, we store $y^k$ for the last 20\% of iterations (i.e., $10^5$ iterations) and calculate their empirical mean, denoted $\bar{y}^n$. The resulting risk budgeting portfolio is then given by $
\bar{u}^n = \frac{\bar{y}^n}{ \|\bar{y}^n\|_1}$, which provides a more accurate estimate by averaging out the fluctuations in the final iterations.

\begin{table}[h]
    \centering
\begin{tabular}{c|cc|cc|cc}
\toprule
{} & \multicolumn{2}{c|}{MAD} & \multicolumn{2}{c|}{Volatility} & \multicolumn{2}{c}{Variantile} \\
$d$ &          SMD &         t-SGD &          SMD &         t-SGD &          SMD &         t-SGD \\
\midrule
10  &  2.26 (0.38) &  2.46 (0.49) &       3.91 (0.96) &        16.30 (6.99) &       4.55 (1.01) &        21.76 (9.14) \\
25  &  1.04 (0.14) &  1.14 (0.17) &       2.05 (0.29) &         2.75 (0.72) &       2.40 (0.36) &         3.32 (0.87) \\
50  &  0.57 (0.05) &  0.61 (0.06) &       1.23 (0.17) &         1.11 (0.17) &       1.40 (0.23) &         1.43 (0.25) \\
100 &  0.30 (0.02) &  0.34 (0.02) &       0.69 (0.07) &         0.65 (0.08) &       0.80 (0.10) &         0.79 (0.11) \\
250 &  0.13 (0.01) &  0.14 (0.01) &       0.30 (0.03) &         0.31 (0.03) &       0.34 (0.04) &         0.35 (0.04) \\

\bottomrule
\end{tabular}
\caption{MDEs of the portfolios computed using the SMD and t-SGD algorithms across different portfolio sizes.}
\label{error_analysis_different_risk_measures}
\end{table}

Table~\ref{error_analysis_different_risk_measures} shows the median of the MDEs with median absolute deviations of the MDEs (in parentheses) across different portfolio sizes $d$ and risk measures over 100 repetitions. The initial choice of the learning rate $\gamma$ is fixed across various portfolio sizes $d$ to avoid tuning the optimal value for each dimension, method, and risk measure. The results indicate that SMD produces more accurate portfolios than t-SGD in most cases, highlighting its suitability for computing risk budgeting portfolios accurately across a range of risk measures. These values, of course, only allow for a comparative analysis, and absolute errors could be further reduced by choosing hyperparameters more carefully.

\section*{Conclusion}

In this article, leveraging the characterization of risk budgeting portfolios as solutions to a strictly convex optimization problem, we proposed and analyzed the convergence of deterministic and stochastic mirror descent (MD) algorithms for computing these portfolios across general risk measures. Due to the unbounded gradient of the value function, standard convergence results for MD algorithms do not directly apply. By developing a tailored tamed version of the gradient, we introduced both deterministic and stochastic MD schemes, establishing their convergence along with a convergence rate for the weighted averaged sequence. The stochastic framework accommodates several commonly used risk measures, such as volatility, expected shortfall, and deviation measures.

Our numerical results illustrate these theoretical findings, demonstrating that both algorithms remain stable as dimensionality increases and perform more effectively than the standard SGD algorithm recently proposed in the literature.

Potential directions for future research include investigating asymptotic error properties by establishing a central limit theorem for the stochastic MD algorithm presented here. Another promising line of inquiry could be to explore MD algorithms when only biased samples $(X^h_n)_{n\geq1}$ of $X$, where $h$ represents the bias parameter, as seen for instance in the two recent works \cite{Atchade:Fort:Moulines} or \cite{costa:gadat:huang}. In this setting, to further manage complexity, one might employ multi-level or multi-step Richardson-Romberg techniques in stochastic approximation schemes, as originally developed in \cite{Fri16} and \cite{Frikha:Huang:MSRR}, and also as applied in recent works on VaR and ES in \cite{Crepey:Frikha:Louzi:1, Crepey:Frikha:Louzi:Pages:2, Crepey:Frikha:Louzi:Spence:3} for the special case of the VaR and ES. It could also be worth investigating the applicability of MD algorithms for calculating risk budgeting portfolios that take into account underlying risk factors, as outlined in \cite{cetingoz2023factor}.

\bibliographystyle{plain}%alpha
\bibliography{article.bib}

\begin{thebibliography}{10}

\bibitem{acerbi2002spectral}
C.~Acerbi.
\newblock Spectral measures of risk: A coherent representation of subjective
  risk aversion.
\newblock {\em Journal of Banking \& Finance}, 26(7):1505--1518, 2002.

\bibitem{AT02}
C.~Acerbi and D.~Tasche.
\newblock On the coherence of {E}xpected {S}hortfall.
\newblock {\em Journal of Banking \& Finance}, 26(7):1487--1503, 2002.

\bibitem{ararat2024mad}
{\c{C}}.~Ararat, F.~Cesarone, M.~{\c{C}}elebi P{\i}nar, and J.~M. Ricci.
\newblock Mad risk parity portfolios.
\newblock {\em Annals of Operations Research}, pages 1--26, 2024.

\bibitem{city33733}
V.~Asimit, L.~Peng, R.~Tunaru, and F.~Zhou.
\newblock Risk budgeting under general risk measures.
\newblock Submitted.

\bibitem{Atchade:Fort:Moulines}
Y.~F. Atchad{{\'e}}, G.~Fort, and R.~Moulines.
\newblock On perturbed proximal gradient algorithms.
\newblock {\em Journal of Machine Learning Research}, 18(10):1--33, 2017.

\bibitem{bai2016least}
X.~Bai, K.~Scheinberg, and R.~Tutuncu.
\newblock Least-squares approach to risk parity in portfolio selection.
\newblock {\em Quantitative Finance}, 16(3):357--376, 2016.

\bibitem{BFP09:2}
O.~Bardou, N.~Frikha, and G.~Pag{\`e}s.
\newblock Recursive computation of value-at-risk and conditional value-at-risk
  using mc and qmc.
\newblock In Pierre L'Ecuyer and Art~B. Owen, editors, {\em Monte Carlo and
  Quasi-Monte Carlo Methods 2008}, pages 193--208, Berlin, Heidelberg, 2009.
  Springer Berlin Heidelberg.

\bibitem{bardou2016cvar}
O.~Bardou, N.~Frikha, and G.~Pag{\`e}s.
\newblock {CVaR} hedging using quantization-based stochastic approximation
  algorithm.
\newblock {\em Mathematical Finance}, 26(1):184--229, 2016.

\bibitem{BFP09:1}
O.~Bardou, N.~Frikha, and G.~Pagès.
\newblock Computing var and cvar using stochastic approximation and adaptive
  unconstrained importance sampling.
\newblock {\em Monte Carlo Methods and Applications}, 15(3):173--210, 2009.

\bibitem{bellini2021risk}
F.~Bellini, F.~Cesarone, C.~Colombo, and F.~Tardella.
\newblock Risk parity with expectiles.
\newblock {\em European journal of operational research}, 291(3):1149--1163,
  2021.

\bibitem{BT07}
A.~Ben-Tal and M.~Teboulle.
\newblock An old-new concept of convex risk measures: the optimized certainty
  equivalent.
\newblock {\em Mathematical Finance}, 17(3):449--476, 2007.

\bibitem{best1991sensitivity}
M.~J. Best and R.~R. Grauer.
\newblock On the sensitivity of mean-variance-efficient portfolios to changes
  in asset means: some analytical and computational results.
\newblock {\em The Review of Financial Studies}, 4(2):315--342, 1991.

\bibitem{boyd2024markowitz}
S.~Boyd, K.~Johansson, R.~Kahn, P.~Schiele, and T.~Schmelzer.
\newblock Markowitz portfolio construction at seventy.
\newblock {\em arXiv preprint arXiv:2401.05080}, 2024.

\bibitem{BR:2012}
B.~Bruder and T.~Roncalli.
\newblock {Managing Risk Exposures Using the Risk Budgeting Approach}.
\newblock {\em Available at SSRN: https://ssrn.com/abstract=2009778}, 2012.

\bibitem{cetingoz2023risk}
A.~R. Cetingoz, J.-D. Fermanian, and O.~Gu{\'e}ant.
\newblock Risk budgeting portfolios: Existence and computation.
\newblock {\em Mathematical Finance}, 2023.

\bibitem{cetingoz2023factor}
A.~R. Cetingoz and O.~Gu{\'e}ant.
\newblock Factor risk budgeting and beyond.
\newblock {\em arXiv preprint arXiv:2312.11132}, 2023.

\bibitem{chaves2012efficient}
D.~Chaves, J.~Hsu, F.~Li, and O.~Shakernia.
\newblock Efficient algorithms for computing risk parity portfolio weights.
\newblock {\em Journal of Investing}, 21(3):150, 2012.

\bibitem{costa:gadat:huang}
M.~Costa, L.~Huang, and S.~Gadat.
\newblock Cv@r penalized portfolio optimization with biased stochastic mirror
  descent.
\newblock {\em forthcoming in Finance \& Stochastics}, 2024.

\bibitem{Crepey:Frikha:Louzi:1}
S.~Cr\'epey, N.~Frikha, and A.~Louzi.
\newblock {A Multilevel Stochastic Approximation Algorithm for Value-at-Risk
  and Expected Shortfall Estimation}.
\newblock {\em forthcoming for Finance \& Stochastics}, 2024.

\bibitem{Crepey:Frikha:Louzi:Pages:2}
S.~Cr\'epey, N.~Frikha, A.~Louzi, and G.~Pag\`es.
\newblock {Asymptotic Error Analysis of Multilevel Stochastic Approximations
  for the Value-at-Risk and Expected Shortfall}.
\newblock {\em {forthcoming for Electronic Journal of Probability}}, 2024.

\bibitem{Crepey:Frikha:Louzi:Spence:3}
S.~Cr\'epey, N.~Frikha, A.~Louzi, and J.~Spence.
\newblock {Adaptive Multilevel Stochastic Approximation of the Value-at-Risk}.
\newblock {\em arXiv:2408.06531}, 2024.

\bibitem{da2023risk}
B.~F.~P. da~Costa, S.~M. Pesenti, and R.~S. Targino.
\newblock Risk budgeting portfolios from simulations.
\newblock {\em European Journal of Operational Research}, 311(3):1040--1056,
  2023.

\bibitem{bayes:risks:21}
P.~Embrechts, T.~Mao, Q.~Wang, and R.~Wang.
\newblock Bayes risk, elicitability, and the expected shortfall.
\newblock {\em Mathematical Finance}, 31(4):1190--1217, 2021.

\bibitem{frikha:sifin:1}
N.~Frikha.
\newblock Shortfall risk minimization in discrete time financial market models.
\newblock {\em {SIAM Journal on Financial Mathematics}}, 5(1):384--414, 2014.

\bibitem{Fri16}
N.~Frikha.
\newblock {Multi-level stochastic approximation algorithms}.
\newblock {\em The Annals of Applied Probability}, 26(2):933 -- 985, 2016.

\bibitem{Frikha:Huang:MSRR}
N.~Frikha and L.~Huang.
\newblock A multi-step richardson–romberg extrapolation method for stochastic
  approximation.
\newblock {\em Stochastic Processes and their Applications},
  125(11):4066--4101, 2015.

\bibitem{FS10}
H.~Föllmer and A.~Schied.
\newblock {\em Convex risk measures}.
\newblock John Wiley \& Sons, Ltd, 2010.

\bibitem{GBRR:2013}
T.~Griveau-Billion, J.-C. Richard, and T.~Roncalli.
\newblock A fast algorithm for computing high-dimensional risk parity
  portfolios.
\newblock 2013.

\bibitem{luan:Nemirovski:Shapiro}
G.~Lan, A.~Nemirovski, and A.~Shapiro.
\newblock Validation analysis of mirror descent stochastic approximation
  method.
\newblock {\em Mathematical Programming}, 134(2):425--458, 2012.

\bibitem{maillard2010properties}
S.~Maillard, T.~Roncalli, and J.~Teiletche.
\newblock The properties of equally weighted risk contribution portfolios.
\newblock {\em The Journal of Portfolio Management}, 36(4):60--70, 2010.

\bibitem{markowitz1952}
H.~Markowitz.
\newblock Portfolio selection.
\newblock {\em The Journal of Finance}, 7(1):77--91, 1952.

\bibitem{markowitz1956optimization}
H.~Markowitz.
\newblock The optimization of a quadratic function subject to linear
  constraints.
\newblock {\em Naval Research Logistics Quarterly}, 3(1-2):111--133, 1956.

\bibitem{mausser2018long}
H.~Mausser and O.~Romanko.
\newblock Long-only equal risk contribution portfolios for cvar under discrete
  distributions.
\newblock {\em Quantitative Finance}, 18(11):1927--1945, 2018.

\bibitem{michaud1989markowitz}
R.~O. Michaud.
\newblock The markowitz optimization enigma: Is ‘optimized’optimal?
\newblock {\em Financial Analysts Journal}, 45(1):31--42, 1989.

\bibitem{NJLS09}
A.~Nemirovski, A.~Juditsky, G.~Lan, and A.~Shapiro.
\newblock Robust stochastic approximation approach to stochastic programming.
\newblock {\em SIAM Journal on Optimization}, 19(4):1574--1609, 2009.

\bibitem{nemirovskij1983problem}
A.~S. Nemirovskij and D.~B. Yudin.
\newblock {\em Problem complexity and method efficiency in optimization}.
\newblock John Wiley and Sons, 1983.

\bibitem{nesterov:07}
Y.~Nesterov.
\newblock Primal-dual subgradient methods for convex problems.
\newblock {\em Mathematical Programming}, 120(1):221--259, 2009.

\bibitem{qian2005risk}
E.~Qian et~al.
\newblock Risk parity portfolios: Efficient portfolios through true
  diversification.
\newblock {\em Panagora Asset Management}, 2005.

\bibitem{RU00}
R.~T. Rockafellar and S.~Uryasev.
\newblock Optimization of {C}onditional {V}alue-at-{R}isk.
\newblock {\em Journal of risk}, 2(3):21--41, 2000.

\bibitem{rockafellar2002conditional}
R.~T. Rockafellar and S.~Uryasev.
\newblock Conditional value-at-risk for general loss distributions.
\newblock {\em Journal of banking \& finance}, 26(7):1443--1471, 2002.

\bibitem{ROCKAFELLAR201333}
R.~T. Rockafellar and S.~Uryasev.
\newblock The fundamental risk quadrangle in risk management, optimization and
  statistical estimation.
\newblock {\em Surveys in Operations Research and Management Science},
  18(1):33--53, 2013.

\bibitem{rock:uryasev:zaba:06}
R.~T. Rockafellar, S.~Uryasev, and M.~Zabarankin.
\newblock Generalized deviations in risk analysis.
\newblock {\em Finance Stochast.}, 10:51--74, 2006.

\bibitem{rock:uryasev:zaba:08}
R.~T. Rockafellar, S.~Uryasev, and M.~Zabarankin.
\newblock Risk tuning with generalized linear regression.
\newblock {\em Mathematics of Operations Research}, 33:712--729, 2008.

\bibitem{roncalli2013introduction}
T.~Roncalli.
\newblock {\em Introduction to Risk Parity and Budgeting}.
\newblock CRC Press, 2013.

\bibitem{S:2013}
F.~Spinu.
\newblock An algorithm for computing risk parity weights.
\newblock Technical report, 2013.

\bibitem{wolfe1959simplex}
P.~Wolfe.
\newblock The simplex method for quadratic programming.
\newblock {\em Econometrica: Journal of the Econometric Society}, pages
  382--398, 1959.

\bibitem{NIPS2017_e6ba70fc}
Z.~Zhou, P.~Mertikopoulos, N.~Bambos, S.~Boyd, and P.~W. Glynn.
\newblock Stochastic mirror descent in variationally coherent optimization
  problems.
\newblock In I.~Guyon, U.~Von Luxburg, S.~Bengio, H.~Wallach, R.~Fergus,
  S.~Vishwanathan, and R.~Garnett, editors, {\em Advances in Neural Information
  Processing Systems}, volume~30. Curran Associates, Inc., 2017.

\end{thebibliography}

%\thispagestyle{empty}
%\sloppy
%\newpage
%\printbibliography[heading=bibintoc]
%\fussy

%\newpage
%\pagenumbering{Roman}
%\pagestyle{appendix}
\appendix

\section{Appendix}\label{sec:appendix}

\subsection{Convergence for the DMD algorithm: proof of Theorem \ref{thm:conv:DMD}}\label{proof:thm:conv:DMD}

\noindent \emph{Step 1: On the convergence of $(y^n)_{n\geq1}$.}

In order to study the convergence of the sequence $(y^k)_{k\geq0}$ towards the unique minimizer $y^{\star}$ of $\Gamma_g$, we recall that Lemma 2.1 in \cite{NJLS09} (combined with \eqref{ineq:strong:convexity}) guarantees that
\begin{equation}\label{key:ineq:bregman:div}
 D_F(w, {\rm P}^m_u(v)) \leq D_F(w, u) + \langle v, w-u \rangle + m \frac{\|v\|_{\infty}^2}{2}
\end{equation}

\noindent for any $w, u \in (\mathbb{R}^{0}_+)^d \cap B_m$, $v \in \mathbb{R}^d$. Selecting $v= \gamma_{k+1} \kappa(y^k) \, \nabla \Gamma_g (y^k)$, $u=y^k$ and $w = y^{\star}$ in the above inequality, and taking $m\geq \| y^{\star}\|_1$, we get
\begin{equation}\label{ineq:rec:DF}
\begin{aligned}
D_F( y^{\star}, y^{k+1}) & \leq D_F( y^{\star}, y^k) - \gamma_{k+1} \langle \kappa(y^k) \nabla \Gamma_g(y^k), y^k-y^{\star} \rangle \\
& \quad + m \gamma_{k+1}^2 \frac{\|\kappa(y^k) \, \nabla \Gamma_g(y^k)\|_{\infty}^2}{2}.
\end{aligned}
\end{equation}

 Assuming that the learning rate satisfies $\sum_{k\geq1} \gamma_k^2 < \infty$ and recalling \eqref{ineq:growth:modified:grad}, the Robbins-Siegmund theorem yields 
\begin{equation}\label{eq:consequence:deterministic:robbins:siegmund:thm}
D_F(y^{\star}, y^{n}) \stackrel{n\rightarrow \infty}{\longrightarrow} D_\infty < \infty \quad  \mbox{ and } \quad  \sum_{k\geq0} \gamma_{k+1} \langle \kappa(y^k) \nabla \Gamma_g(y^k), y^k-y^{\star} \rangle < \infty.
\end{equation}

Since $\sum_{k\geq1} \gamma_k = \infty$, we deduce that $\lim\inf_{n \to \infty } \langle \kappa(y^n) \nabla \Gamma_g(y^n), y^n - y^{\star} \rangle =0 $. The boundedness of the sequence $(y^n)_{n\geq0}$ guarantees the existence of a subsequence $(\varphi(n))_{n\geq0}$ such that
$$
y^{\varphi(n)}\rightarrow y^\infty \in (\mathbb{R}_+)^d \cap B_m, \quad \mbox{ and } \quad \langle \kappa(y^{\varphi(n)}) \nabla \Gamma_{g}(y^{\varphi(n)}), y^{\varphi(n)}-y^{\star} \rangle =0.
$$

By continuity of $\mathbb{R}_+^d \ni y\mapsto \kappa(y) \nabla   \Gamma_g(y)$, it follows that $\lim_{n \to \infty} \langle \kappa(y^{\varphi(n)}) \nabla \Gamma_g(y^{\varphi(n)}), y^{\varphi(n)}-y^{\star} \rangle = \langle \kappa(y^{\infty})  \nabla \Gamma_g(y^{\infty}), y^{\infty}-y^{\star} \rangle = 0$ which in turn, by Lemma \ref{lem:mean:reverting}, implies that $y^{\infty} = y^{\star}$. 

Coming back to \eqref{eq:consequence:deterministic:robbins:siegmund:thm}, we eventually deduce that
$$
\lim_{n \to \infty} D_F(y^{\star}, y^n) = \lim_{n \to \infty} D_F( y^{\star}, y^{\varphi(n)}) = 0 
$$

\noindent which clearly implies that $(y^n)_{n\geq1}$ converges towards $y^{\star}$.

\noindent \emph{Step 2: On the convergence rate of $(\bar{y}^n)_{n\geq1}$.}

In order to derive the corresponding convergence rate, we come back to \eqref{ineq:rec:DF} and write
$$
 \gamma_{k+1} \langle \kappa(y^k)  \nabla \Gamma_g(y^k), y^k-y^{\star} \rangle \leq D_F( y^{\star}, y^k) - D_F(y^{\star}, y^{k+1}) + \frac12 m \gamma_{k+1}^2 M_{\star}^2.
$$

Summing over $k=0, \cdots, n-1$ the previous inequality and recalling that $D_F(y^{k+1}, y^{\star})\geq 0$, we get
$$
\sum_{k=1}^{n} \gamma_k \langle\kappa(y^{k-1}) \, \nabla \Gamma_g(y^{k-1}), y^{k-1}-y^{\star} \rangle \leq D_F(y^{\star}, y^{0}) + \frac12 m M_\star^2 \sum_{k=1}^{n} \gamma_{k}^2.
$$

By convexity of $\Gamma_g$ and Jensen's inequality, we obtain
\begin{align*}   
\sum_{k=1}^{n} \gamma_k \kappa(y^{k-1}) \langle \nabla \Gamma_g(y^{k-1}), y^{k-1}-y^{\star} \rangle & \geq (\min_{0\leq k \leq n-1} \underline{y}^k \wedge 1) \sum_{k=1}^{n} \gamma_k  \langle \nabla \Gamma_g(y^{k-1}), y^{k-1}-y^{\star} \rangle\\
& \geq  (\min_{0\leq k \leq n-1} \underline{y}^k \wedge 1 ) \sum_{k=1}^n \gamma_k ( \Gamma_g(y^{k-1}) - \Gamma_g(y^{\star}) ) \\
& \geq (\min_{0\leq k \leq n-1} \underline{y}^k \wedge 1 ) (\sum_{k=1}^{n} \gamma_k) ( \Gamma_g(\bar{y}^{n}) - \Gamma_g(y^{\star}) ) 
\end{align*}

\noindent recalling that $(\bar{y}^{n})_{n\geq1}$ is the averaged version of $(y^n)_{n \geq 0}$. Combining the two previous inequalities eventually yields
$$
\Gamma_g(\bar{y}^{n}) - \Gamma_g(y^{\star}) \leq \frac{D_F(y^{\star} ,y^0) + \frac12 m M^2_{\star} \sum^n_{k=1} \gamma_k^2}{(\min_{0\leq k \leq n-1} \underline{y}^k \wedge 1 ) \sum^n_{k=1} \gamma_k}, \quad n\geq1,
$$

\noindent which concludes the proof.

\subsection{Proof of Proposition \ref{prop:rb:problem:stochastic:framework}}\label{subsec:prop:rb:problem:stochastic:framework}

\noindent \emph{Step 1:} The regularity and integrability conditions combined with the dominated convergence theorem guarantee that $h$ is continuously differentiable on $\mathbb{R}\times (\mathbb{R}_+^{0})^d$. The expression for the derivatives $\partial_\xi h(\xi, y)$ and $\partial_{y_i} h(\xi, y)$, $i=1,\cdots, d$, is easily obtained by differentiating under expectation. The convexity of $h$ then guarantees that $\Argmin h = \left\{ \nabla h = 0\right\}$.

\noindent\emph{Step 2:} We here prove that the map $(\mathbb{R}^0_+)^d\ni y\mapsto g(r_\rho(y))$ defined by \eqref{def:g:r:rho:stochastic:case} is continuously differentiable with a derivative given by $\nabla (g(r_\rho(y))) = \mathbb{E}[-X \partial_x L(\xi^{\star}(y),-\langle y, X\rangle)]$. Let $ y, h \in \mathbb{R}^d$. Note that the minimizers $\xi^{\star}(y+h)$, $\xi^{\star}(y)$ of $\mathbb{E}[L(., -\langle (y+h), X\rangle)]$ and $\mathbb{E}[L(., -\langle y, X\rangle)]$ respectively, satisfy
\begin{equation*}
\begin{aligned}
g(r_\rho(y+h)) & - g(r_\rho(y)) \\
& = \mathbb{E}[L(\xi^{\star}(y+h), -\langle(y+h),X\rangle)] - \mathbb{E}[L(\xi^{\star}(y), -\langle y, X\rangle)] \\
& \leq \mathbb{E}[L(\xi^{\star}(y), -\langle (y+h), X\rangle)] - \mathbb{E}[L(\xi^{\star}(y), -\langle y,X\rangle)]\\
& =  \langle \mathbb{E}[-X \partial_x L(\xi, -\langle y,X\rangle)]_{\xi = \xi^{\star}(y)}, h \rangle + o(\|h\|)
\end{aligned}
\end{equation*}
\noindent as $\|h\| \downarrow 0$. Similarly, one has
\begin{equation*}
\begin{aligned}
& g(r_\rho(y+h))  - g(r_\rho(y)) \\
& \geq \mathbb{E}[L(\xi^{\star}(y+h), -\langle (y+h),X\rangle)] - \mathbb{E}[L(\xi^{\star}(y+h), -\langle y, X\rangle)] \\
& =  \langle \mathbb{E}[-X \partial_x L(\xi, -\langle y, X\rangle)]_{\xi = \xi^{\star}(y+h)}, h \rangle \\
& \quad + \int_0^1 \langle \mathbb{E}[-X\partial_x L(\xi, -\langle (y + t h),X\rangle)]_{\xi = \xi^{\star}(y+h)} - \mathbb{E}[-X \partial_x L(\xi, -\langle y, X\rangle)]_{\xi = \xi^{\star}(y+h)}, h \rangle \ dt \\
& = \langle \mathbb{E}[-X \partial_x L(\xi, -\langle y,X\rangle)]_{\xi = \xi^{\star}(y)}, h \rangle  \\
& \quad + \langle \mathbb{E}[-X \partial_x L(\xi, -\langle y, X\rangle)]_{\xi = \xi^{\star}(y+h)} - \mathbb{E}[-X\partial_x L(\xi, -\langle y,X\rangle)]_{\xi = \xi^{\star}(y)}, h \rangle \\
& \quad + \int_0^1 \langle \mathbb{E}[-X \partial_x L(\xi, -\langle (y + t h), X\rangle)]_{\xi = \xi^{\star}(y+h)} - \mathbb{E}[-X \partial_x L(\xi, -\langle y, X\rangle)]_{\xi = \xi^{\star}(y+h)}, h \rangle \ dt.
\end{aligned}
\end{equation*}
Our aim now is to prove that the two last terms appearing in the right-hand side of the above inequality are $o(\|h\|)$. Clearly, it suffices to prove that 
\begin{equation}\label{first:continuity:res:proof:prop}
\lim_{h\downarrow 0} \mathbb{E}[X \partial_x L(\xi, -\langle y,X\rangle)]_{\xi = \xi^{\star}(y+h)}= \mathbb{E}[X \partial_x L(\xi, -\langle y,X\rangle)]_{\xi = \xi^{\star}(y)}
\end{equation}
\noindent and that 
\begin{equation}\label{second:continuity:res:proof:prop}
\lim_{h\downarrow 0}\mathbb{E}[X \partial_x L(\xi, -\langle (y + t h), X\rangle)] = \mathbb{E}[X \partial_x L(\xi, -\langle y,X\rangle)]
\end{equation}
\noindent locally uniformly in $\xi$.

Note that the convexity of the maps $\xi \mapsto \mathbb{E}[L(\xi, -\langle y, X\rangle)]$, and $\xi \mapsto \mathbb{E}[L(\xi, -\langle (y+h),X\rangle)]$ yields {\small $(\xi -\xi^\star(y))\mathbb{E}[\partial_\xi L(\xi, -\langle y, X\rangle)]>0$} for any $\xi \neq \xi^{\star}(y)$ and $(\xi -\xi^\star(y+h))\mathbb{E}[\partial_\xi L(\xi, -\langle (y+h),X\rangle)]>0$ for any $\xi \neq \xi^{\star}(y+h)$. Moreover, the continuity of $\mathbb{R} \times (\mathbb{R}^0_+)^d \ni (\xi, y) \mapsto \mathbb{E}[\partial_\xi L(\xi,-\langle y,X\rangle)]$ guarantees that for any fixed $y$, $\xi \mapsto \mathbb{E}[\partial_\xi L(\xi,-\langle (y+h),X\rangle)]$ converges to $\xi \mathbb{E}[\partial_\xi L(\xi,-\langle y,X\rangle)]$ as $h\downarrow 0$ locally uniformly. Now, \cite[Theorem~2.6]{Fri16} guarantees that $\xi^{\star}(y+h) \rightarrow \xi^{\star}(y)$ as $\|h\| \downarrow 0$ which combined with the continuity of $\xi \mapsto \mathbb{E}[X \partial_z L(\xi, -\langle y,X\rangle)]$ implies \eqref{first:continuity:res:proof:prop}.

The second point \eqref{second:continuity:res:proof:prop} is a consequence of the continuity of $\mathbb{R} \times \mathbb{R}^d \ni (\xi, y)\mapsto \mathbb{E}[X\partial_x L(\xi, -\langle y, X\rangle)]$.

Combining the above arguments, we conclude that $y\mapsto g(r_\rho(y))$ is continuously differentiable and satisfies $\nabla (g(r_\rho(y))) = \mathbb{E}[-X \partial_x L(\xi^{\star}(y),-\langle y,X\rangle)]$.

\noindent \emph{Step 3: } If $(\xi^\star, y^\star) \in \Argmin h$, then $\partial_{y_i} h(\xi^\star, y^{\star}) = 0$, $i=1, \cdots, d$, which, according to the conclusion of the previous step, implies that
$$
\partial_{y_i} \Gamma_g(y^\star)=\partial_{y_i} (g(r_\rho(y^{\star}))) - \frac{b_i}{y^{\star}_i} = 0, \quad i=1, \cdots, d.
$$

Hence, $y^\star$ is the unique minimizer of $\Gamma_g$. The proof is now complete.

\subsection{Convergence of the SMD algorithm: proof of Theorem \ref{thm:conv:SMD:algorithm}}\label{subsec:thm:conv:SMD:algorithm}

\noindent \emph{Step 1: On the convergence of $(z^n)_{n\geq1}$} 

We use the inequality \eqref{key:ineq:bregman:div} with $G$ instead of $F$ (noting \eqref{ineq:bregman:div:G}) namely
$$
D_G(z, {\rm P}_{z'}^m(v)) \leq D_{G}(z, z') + \langle v, z-z' \rangle + m\vee 1 \|v\|_\infty^2.
$$

\noindent with $v= \gamma_{k+1} (\partial_\xi H(z^k, X^{k+1}), \kappa(y^k)  \nabla_y H(z^k, X^{k+1}))^{T}$, $z'=z^k$ and $z = z^{\star}$. Hence,
\begin{equation} \label{ineq:rec:stochastic:MD}
\begin{aligned}
D_G(z^\star, z^{k+1})  \leq \  & D_G(z^\star, z^k)  - \gamma_{k+1} \left\langle  \mathbb{E}\left[\begin{pmatrix} \partial_{\xi} H(z, X) \\ \kappa(y) \nabla_{y}H(z, X) \end{pmatrix} \right]_{|z=z^{k}}, z^k-z^\star \right\rangle  \\
& - \gamma_{k+1} \langle \Delta^{k+1}, z^k-z^\star \rangle  +  m\vee 1 \gamma_{k+1}^2 Y_{k+1},
\end{aligned}
\end{equation}

\noindent where we used the fact that $X^{k+1}$ is independent of $(z^{j})_{0\leq j \leq k}$ and introduced the notations
\begin{equation}
\begin{aligned}
\Delta^{k+1} & = \begin{pmatrix} \partial_{\xi} H(z^k, X^{k+1}) \\ \kappa(y^k) \nabla_{y} H(z^{k}, X^{k+1}) \end{pmatrix} -  \mathbb{E}\left[\begin{pmatrix} \partial_{\xi} H(z, X) \\ \kappa(y) \nabla_{y}H(z, X) \end{pmatrix} \right]_{|z=z^{k}},\\
Y_{k+1} & := \left\| \begin{pmatrix} \partial_{\xi} H(z^k, X^{k+1}) \\ \kappa(y^k) \nabla_{y} H(z^{k}, X^{k+1}) \end{pmatrix}\right\|_{\infty}^2. 
\end{aligned}
\quad \quad k \geq0.
\end{equation}

 Note that since $z^k$ is $\mathcal{F}_k$-measurable and $X^{k+1}$ is independent of $\mathcal{F}_k$, $(\Delta^k)_{k\geq1}$ is a sequence of $\mathcal{F}$-martingale increments
$$
\mathbb{E}[\Delta^{k+1}|\mathcal{F}_k] = 0, \quad \mbox{for } k\geq0.
$$

Now, the previous identity together with \eqref{ineq:rec:stochastic:MD} and
observing that 
$$
\mathbb{E}[Y_{k+1}|\mathcal{F}_k] \leq N^2_{\star}:=\sup_{(\xi,y) \in \mathbb{R} \times \mathbb{R}_+^d  \cap B_m)} \mathbb{E}\Big[\Big\|\ \begin{pmatrix} \partial_{\xi} H(z, X) \\ \kappa(y) \nabla_{y} H(z, X) \end{pmatrix}  \Big\|^2_{\infty}\Big] < \infty, \quad k\geq0
$$
\noindent yields
\begin{equation} \label{ineq:bregman:divergence:zk+1:to:zk}
\begin{aligned}
\mathbb{E}[D_G(z^\star, z^{k+1})|\mathcal{F}_k]  \leq \  & D_G(z^\star, z^k)  - \gamma_{k+1} \left\langle  \mathbb{E}\left[\begin{pmatrix} \partial_{\xi} H(z, X) \\ \kappa(y) \nabla_{y}H(z, X) \end{pmatrix} \right]_{|z=z^{k}}, z^k-z^\star \right\rangle  \\
&  + m\vee 1 N^2_\star \gamma_{k+1}^2.
\end{aligned}
\end{equation}

The Robbins-Siegmund theorem guarantees that
\begin{equation}\label{eq:consequence:stochastic:robbins:siegmund:thm}
D_G(z^\star, z^n) \stackrel{n\rightarrow \infty}{\longrightarrow} D_\infty \in L^{1}(\mathbb{P}) \quad a.s.
\end{equation}
\noindent and 
$$
\sum_{n\geq0} \gamma_{n+1}  \left\langle  \mathbb{E}\left[\begin{pmatrix} \partial_{\xi} H(z, X) \\ \kappa(y) \nabla_{y}H(z, X) \end{pmatrix} \right]_{|z=z^{n}}, z^n-z^\star \right\rangle < \infty, \quad a.s. 
$$

The first assertion above implies that the sequence $(z^n)_{n\geq0}$ is bounded while the second together with Lemma \ref{mean:reverting:stochastic} and the condition $\sum_{k\geq1} \gamma_k = \infty$ guarantees that 
$$
\lim\inf_{n \rightarrow \infty} \left\langle  \mathbb{E}\left[\begin{pmatrix} \partial_{\xi} H(z, X) \\ \kappa(y) \nabla_{y}H(z, X) \end{pmatrix} \right]_{|z=z^{n}}, z^n-z^\star\right\rangle =0.
$$

As a consequence, we may assume the existence of a subsequence $(\varphi(n))_{n\geq0}$ such that
$$
z^{\varphi(n)}\rightarrow z^\infty \in \mathbb{R} \times (\mathbb{R}_+^d \cap B^d_m)
$$
\noindent and
$$
\lim_{n \rightarrow \infty}  \left\langle  \mathbb{E}\left[\begin{pmatrix} \partial_{\xi} H(z, X) \\ \kappa(y) \nabla_{y}H(z, X) \end{pmatrix} \right]_{|z=z^{\varphi(n)}}, z^{\varphi(n)}-z^\star \right\rangle =0.
$$
By continuity of the map $z\mapsto (\mathbb{E}[\partial_{\xi} H(z, X)], \, \mathbb{E}[\kappa(y) \nabla_{y}H(z, X)])$, it follows that:
\begin{align*}
\lim_{n \rightarrow \infty}  & \left\langle   \mathbb{E}\left[\begin{pmatrix} \partial_{\xi} H(z, X) \\ \kappa(y) \nabla_{y}H(z, X) \end{pmatrix} \right]_{|z=z^{\varphi(n)}}, z^{\varphi(n)}-z^\star \right\rangle \\
& = 
 \left\langle \mathbb{E}\left[\begin{pmatrix} \partial_{\xi} H(z^\infty, X) \\ \kappa(y^\infty) \nabla_{y}H(z^\infty, X) \end{pmatrix} \right] , z^{\infty}-z^\star \right\rangle = 0
\end{align*}
and, by Lemma \ref{mean:reverting:stochastic}, implies that $z^\infty = z^{\star}$. Subsequently, coming back to \eqref{eq:consequence:stochastic:robbins:siegmund:thm}, we deduce that
$$
\lim_{n \rightarrow \infty} D_G(z^\star, z^n) = \lim_{n \rightarrow \infty} D_G(z^{\varphi(n)} , z^{\star}) = 0 
$$

\noindent which clearly implies that the sequence $(z^n)_{n\geq0}$ converges $a.s.$ towards $z^{\star}$. Note additionally that taking expectation on both sides of \eqref{ineq:bregman:divergence:zk+1:to:zk} and using Lemma \ref{mean:reverting:stochastic} together with a direct induction argument give
\begin{equation}\label{uniform:bound:bregman:divergence:z^n}
\sup_{n\geq1}\mathbb{E}[D_G(z^{n} , z^{\star})]\leq \mathbb{E}[D_G(z^\star, z^0)] + m\vee 1 M_\star^2 \sum_{n\geq 1} \gamma_n^2.
\end{equation}

The above bound will be useful in order to investigate the convergence rate of $(\bar{z}^n)_{n\geq1}$ in the next step.

\noindent \emph{Step 2: On the convergence rate of $(\bar{z}^n)_{n\geq 1}$}

 We now come back to \eqref{ineq:rec:stochastic:MD} getting 
\begin{equation*} 
\begin{aligned}
\gamma_{k+1} & \left\langle  \mathbb{E}\left[\begin{pmatrix} \partial_{\xi} H(z, X) \\ \kappa(y) \nabla_{y}H(z, X) \end{pmatrix} \right]_{|z=z^{k}}, z^k-z^\star \right\rangle \\
\quad &  \leq D_G(z^\star, z^k)  - D_G(z^\star,z^{k+1}) - \gamma_{k+1} \langle \Delta^{k+1}, z^k-z^\star \rangle  
 \\ &  \quad + m \vee 1 \gamma_{k+1}^2 Y_{k+1}
\end{aligned}
\end{equation*}

Summing over $k=0,1, \cdots, n$ the previous inequality and recalling that $D_G(z^{\star}, z^{n+1}) \geq  0$, we get

\begin{equation}\label{first:ineq:conv:bar:zn}
\begin{aligned}
\sum_{k=0}^{n} \gamma_{k+1} & \Big\langle  \mathbb{E}\left[\begin{pmatrix} \partial_{\xi} H(z, X) \\ \kappa(y) \nabla_{y}H(z, X) \end{pmatrix} \right]_{|z=z^{k}} , z^k-z^\star \Big\rangle \\
& \leq  D_G(z^\star, z^0) - M_n  +  m\vee 1 \sum_{k=0}^{n} \gamma_{k+1}^2 Y_{k+1}
\end{aligned}
\end{equation}

\noindent where we introduced the notation
$$
M_n := \sum_{k=0}^n  \gamma_{k+1} \langle \Delta^{k+1}, z^k-z^\star \rangle, \quad n\geq0.
$$

Note that since $(\Delta^k)_{k\geq1}$ is a sequence of $\mathcal{F}$-martingale increments, $(M_n)_{n\geq0}$ is an $\mathcal{F}$-martingale.

Combining the convexity of $h$ with Jensen's inequality, we obtain

\begin{equation}\label{second:ineq:conv:bar:zn}
\begin{aligned}
 \sum_{k=0}^{n} \gamma_{k+1}&  \left\langle  \mathbb{E}\left[\begin{pmatrix} \partial_{\xi} H(z, X) \\ \kappa(y) \nabla_{y}H(z, X) \end{pmatrix} \right]_{|z=z^{k}}, z^k-z^\star \right\rangle    \\
 & \quad \geq \sum_{k=0}^{n} \gamma_{k+1}  \kappa(y^k) \langle \nabla h(z^k) , z^{k} - z^{\star}  \rangle   \\ 
 & \quad \geq  \sum_{k=0}^{n} \gamma_{k+1} \kappa(y^k)  (h(z^k) - h(z^{\star})) \\
 & \quad \geq \min_{0\leq k\leq n}\kappa(y^k)   (\sum_{k=0}^{n} \gamma_{k+1}) (h(\bar{z}^n) - h(z^{\star})).
\end{aligned}
\end{equation}

Combining \eqref{first:ineq:conv:bar:zn} with \eqref{second:ineq:conv:bar:zn} eventually yields
\begin{equation}\label{upper:bound:diff:h:bar:zn:h:z:star}
\begin{aligned}
h(\bar{z}^n)- h(z^*) \leq & \frac{1}{\min_{0\leq k\leq n}\kappa(y^k)}  \frac{1}{\sum_{k=0}^{n} \gamma_{k+1} } \Bigg[ D_G(z^*, z^0) - M_n  + m\vee 1  \sum_{k=0}^{n} \gamma_{k+1}^2 Y_{k+1} \Bigg], \quad a.s.
\end{aligned}
\end{equation}

\noindent \emph{Step 3: }

To conclude the proof, we provide some uniform controls on the two last terms involved in the right-hand side of the previous upper-bound \eqref{upper:bound:diff:h:bar:zn:h:z:star}. Recall that $\left\{ M_n := \sum_{k=0}^n \gamma_{k+1} \langle \Delta^{k+1}, z^k-z^\star \rangle, n\geq1 \right\}$ is an $\mathcal{F}$-martingale. Since
$$
\mathbb{E}[\|\Delta^{k+1}\|_2^2 | \mathcal{F}_{k} ] \leq d \mathbb{E}\Big[\Big\|\ \begin{pmatrix} \partial_{\xi} H(z, X) \\ \kappa(y) \nabla_{y} H(z, X) \end{pmatrix}  \Big\|^2_{\infty}\Big] \leq d  N_\star^2
$$
\noindent we get
$$
\langle M \rangle_n \leq d M_\star^2 \sup_{k\geq0}\|z^k-z^\star\|^2_2 \sum_{k=0}^{n}\gamma_{k+1}^2 
$$
\noindent which in turn implies that $\langle M\rangle_\infty = \lim_n \langle M \rangle_n <\infty$ $a.s.$ Hence, $(M_n)_{n\geq1}$ converges $a.s.$ to $M_\infty< \infty$ $a.s.$ Moreover, from \eqref{ineq:bregman:div:G} and \eqref{uniform:bound:bregman:divergence:z^n}, we deduce 
$$
\sup_{n\geq0} \mathbb{E}[\|z^\star -z^n\|_2^2]\leq 2(m\vee 1) \sup_{n\geq 0}\mathbb{E}[D_G(z^\star, z^n)] \leq \mathbb{E}[D_G(z^\star, z^0)] + \frac12 (m\vee 1) N_\star^2 \sum_{n\geq 1} \gamma_n^2< \infty
$$

\noindent so that
$$
\sup_{n\geq1}\mathbb{E}[M_n^2]\leq \sum_{n\geq0}\gamma_{n+1}^2 \mathbb{E}[\|\Delta^{n+1}\|^2_2 \|z^\star-z^n\|_2^2]\leq d N_\star^2 \sup_{n\geq0} \mathbb{E}[\|z^\star -z^n\|_2^2] \sum_{n\geq0} \gamma_{n+1}^2<\infty. 
$$

 Hence, $(M_n)_{\geq 1}$ is bounded in $L^2(\mathbb{P})$.

To conclude, note that the third term in \eqref{upper:bound:diff:h:bar:zn:h:z:star} is bounded $a.s.$ and in $L^{1}(\mathbb{P})$ since 
$$
\mathbb{E}\Big[ \sum_{n\geq0} \gamma_{n+1}^2 Y_{k+1} \Big] \leq N_\star^2 \sum_{n\geq0} \gamma_{n+1}^2.
$$
The proof is now complete.

%\printbibliography
\end{document}